\documentclass[useAMS,usenatbib,graphicx]{mn2e}

\usepackage{graphicx}

\title[Striations, integrals, hourglasses and collapse]{Striations, integrals, hourglasses and collapse - thermal instability driven magnetic simulations of molecular clouds}
\author[C. J. Wareing et al.]{C. J. Wareing$^{1}$\thanks{E-mail:
C.J.Wareing@leeds.ac.uk}, J. M. Pittard$^{1}$ and S. A. E. G. Falle$^{2}$\\
$^{1}$School of Physics and Astronomy, University of Leeds, Leeds, LS2 9JT, U.K.\\
$^{2}$School of Mathematics, University of Leeds, Leeds, LS2 9JT, U.K.}

\begin{document}

\date{Accepted 2020 November 2. Received 2020 November 2; in original form 2020 June 4}

\pagerange{\pageref{firstpage}--\pageref{lastpage}} \pubyear{2002}

\maketitle

\label{firstpage}

\begin{abstract}
The MHD version of the adaptive mesh refinement (AMR) code, MG, 
has been employed to study the interaction of thermal instability, magnetic fields
and gravity through 3D simulations of the formation of collapsing cold clumps 
on the scale of a few parsecs, inside a larger molecular cloud. The diffuse atomic initial 
condition consists of a stationary, thermally unstable, spherical cloud 
in pressure equilibrium with lower density surroundings and threaded by a uniform magnetic 
field. This cloud was seeded with 10\% density perturbations 
at the finest initial grid level around n=1.1\,cm$^{-3}$ and evolved
with self-gravity included from the outset. Several cloud diameters were considered (100\,pc, 
200\,pc and 400\,pc) equating to several cloud masses (17,000\,M$_\odot$, 136,000\,M$_\odot$
and $1.1\times10^6$\,M$_\odot$). Low-density magnetic-field-aligned striations were observed as the clouds
collapse along the field lines into disc-like structures. The induced flow along field lines leads to
oscillations of the sheet about the gravitational minimum and an integral-shaped appearance. 
When magnetically supercritical, the clouds then collapse and generate hourglass magnetic field 
configurations with strongly intensified magnetic fields, reproducing observational behaviour.
Resimulation of a region of the highest mass cloud at higher resolution  
forms gravitationally-bound collapsing clumps within the sheet that contain clump-frame
supersonic ($M\sim5$) and super-Alfv\'enic ($M_{\rm A}\sim4$) velocities. Observationally
realistic density and velocity power spectra of the cloud and densest clump are obtained.
Future work will use these realistic initial conditions to study individual star and cluster 
feedback.
\end{abstract}

	
\section{Introduction}\label{intro}

Magnetic fields are ubiquitous in star formation across all scales.
Filamentary structure, such as that revealed through
  the {\it Herschel} satellite \cite[see, for example,
  Section 2 of the review of][ and references therein]{andre14}
has been shown to be threaded by these magnetic fields
by observations with such instruments as
  {\it POL-2} with {\it SCUBA-2} on the {\it JCMT} \citep{pattle19}
and most recently {\it ALMA}, amongst other interferometers \citep{hull19}. The effects of
  magnetic fields are seen through their interaction with other
  physical processes (e.g. gravity and turbulence) during both the
  formation of molecular clouds and when feedback processes (radiative
  and mechanical) start, once stars form. These interactions result in
  the formation of such structure as striations and integral-shaped
  filaments and complex magnetic field morphologies, e.g. hourglass
  configurations. Such features provide clues as to the wider importance of
  magnetic fields in the star formation process, which is still debated \citep{hennebelle19}.
For instance, some
numerical results imply that magnetic fields are minor players
in setting either the star formation rate or the initial mass function
\citep{ntor19,krumholz19}. In contrast, models such as that of global hierarchial collapse \citep{vazquez19} emphasize the
importance of magnetic fields throughout. The recent
research topic regarding ``the role of magnetic fields in the formation of stars"
provides an in-depth review for the interested reader \citep{ward20}.
In the following we review some of the observed features which motivate our current work.

Striations - elongated structures in the low column density parts of molecular clouds \citep{goldsmith08}
- appear as separate structures in the Taurus molecular cloud \citep{palm13} and in the Polaris flare
\citep{pano15}. They also appear to connect to the denser filaments.
They have previously been interpreted as streamlines along which material flows toward (or away from) more
dense filaments or clumps \citep{cox16}. \cite{li13} concluded that strong magnetic fields could stabilize guiding channels
of sub-Alfv{\'e}nic flows toward dense filaments, forming striations as a natural result of gravitational contraction,
while \cite{malinen16} concluded that striations were in close alignment
with the magnetic field.

\cite{tritsis16} investigated how striations form through 2D and 3D numerical studies, employing ideal
MHD simulations. They adopted four 2D models, creating striations through (1) sub-Alfv{\'e}nic flow along field lines,
(2) super-Alfv{\'e}nic flow along field lines, (3) sub-Alfv{\'e}nic flow perpendicular to field lines through the
Kelvin-Helmholtz instability, and (4) nonlinear coupling of MHD waves due to density inhomogeneities.
Initial conditions were based on observational estimates \citep{goldsmith08}: a conservative magnetic field
strength of 15$\mu$G and a background number density of n$=200\pm10$\,cm$^{-3}$. A constant temperature
of 15\,K was adopted for all their models, resulting in a sound speed of $\sim0.35\,$km\,s$^{-1}$, an Alfv\'en speed of $\sim1.58\,$km\,s$^{-1}$
and a plasma $\beta$ parameter $\sim 0.1$. They determined that the
first three models do not reproduce
the density contrast inferred from observations. They found a maximum possible contrast in the simulations of isothermal flows (models 1-3) of
0.03\%, compared to an observed mean contrast of $\sim$25\%. Nonlinear coupling of MHD waves 
was able to produce a contrast up to 7\%, a factor of $\sim3$ times smaller than observations even in their 3D simulations, but was
adopted as the most probable formation mechanism of striations. Finally, the authors noted that elongated structures
observed at high Galactic latitudes in the diffuse interstellar
medium (usually referred to as fibers) are similarly
well-ordered with respect to magnetic fields and thus postulate that striations and fibres may share a common
formation mechanism. However, we emphasize that the thermal conditions of the ISM are considerably different and thus
isothermal 15\,K simulations based on molecular cloud conditions cannot establish such a connection.

Turning to magnetic field configurations, recent work from the BISTRO collaboration 
\citep{pattle17,liu19b,wang19,coude19,doi20} has highlighted
how the magnetic field, often found to be fairly uniform in low-to-medium density surroundings, has an hourglass 
morphology around areas of high density \citep{sch98,girart09}. This can be understood in
terms of magnetically supercritical conditions where the effect of
gravity is now dominant. \cite{crutcher12} elucidated this in their plot of magnetic field strength versus density
derived from observations. The fit to data shows that the magnetic field strength remains fairly constant until a critical 
density is reached, found by Crutcher to be on the order of a few hundred particles per cubic centimetre, and then 
increases linearly with density. \cite{pattle17} noted the hourglass morphology of the magnetic field in the OMC 1 
region of the Orion A filament and derived a magnetic field of magnitude $6.6 \pm 4.7$ mG, three orders of magnitude 
greater than the typical background magnetic field of around 10$\mu$G. Further BISTRO survey observations revealed 
measurements of $0.63\pm0.41$\,mG in the Oph-B2 sub-clump \citep{soam18} and $0.5\pm0.2$\,mG toward the
central hub of the IC5146 filamentary cloud \citep{wang19}\footnote{For a more complete discussion of the advantages
and disadvantages of the technique used to derive magnetic field strengths, which may over-estimate field
strengths due to unresolved structure, the interested reader is referred to the recent review by \cite{pattle19},
where observations of high magnetic field strengths are discussed.}.
It is generally accepted that an hourglass morphology is a natural consequence of collapse under gravity. The change
from typically magnetically sub-critical, low-density initial
conditions to magnetically supercritical collapse is accompanied by large
increases in the magnetic field strength. However, such large changes remain difficult to reproduce in simulations, as discussed in a recent
review of numerical methods for simulating star formation \citep{teyssier19}.

The Integral Shaped Filament (ISF) in the Orion A molecular cloud \citep{bally87,stutz16} has been proposed by
\cite{stutz16} to have formed in a slingshot model, where the gas is undergoing oscillations driven by the
interaction between gravity and the magnetic field. The key observed radial velocity gradients they derived
were later confirmed independently by \cite{kong18}, who proposed that the morphology of their position-velocity
diagrams was consistent with this wave-like perturbation in the ISF gas. \cite{stutz18} more recently compared Gaia
parallaxes of young stars in the region to the gas velocities and concluded the ISF has properties consistent with
a standing wave. Most recently, \cite{lobos19} detected velocity gradients on large scales
that are consistent with this wave-like structure, though they also detected small-scale twisting and turning 
structures superimposed on this large-scale structure, leading them to conclude that the structure is even more
complex than previously appreciated. Again, they reinforce the view that the interaction of the magnetic field
and gravity are key to understanding the ISF.

Concurrently, \cite{tritsis18} revealed a `hidden' dimension of the Musca molecular cloud, via the first application of
magnetic seismology. Specifically, they report the detection of normal vibrational modes in the
isolated Musca cloud, allowing the determination of the 3D nature of the cloud. They conclude that Musca is vibrating 
globally, with the characteristic modes of a sheet viewed edge on, not a filament. This model
is dependent on their favoured theoretical explanation of striations involving the excitation of fast magnetosonic
waves compressing the gas and forming ordered structure parallel to the magnetic field. We showed similar 
sheet-like structure forming in our first paper \citep{wareing16} by different thermal means. Whilst not discussed at that time,
the natural formation of the sheet as the MHD cloud collapses along field-lines, with increasing velocity toward the
centre of the gravitational potential, leads to a period during which
there is an oscillation of the sheet about the gravitational 
potential minimum, as the thermal flow along the magnetic field lines
settles to the gravitational potential minimum of the resulting cloud.

To better understand these observations and issues, we study here the 
thermal-gravitational-magnetic interactions that occur during the formation of molecular clouds from 
thermally unstable origins and examine how they can lead to the formation of such structure 
and field configurations, as well as realistic collapsing clumps that will form stars.
We concentrate on the interplay of the thermal instability \citep{parker53,field65} with
the effects of magnetic fields and self-gravity. To keep our study as
straightforward as possible we begin with a low-density cloud of quiescent 
diffuse medium initially in the thermally unstable phase. There is no
initial flow. A uniform magnetic field threads the cloud, which is in
pressure equilibrium with its lower-density surroundings. The simulations
 include accurate thermodynamics, self-gravity and magnetic fields. The aim is to discern whether
thermal instability alone can create structures with high enough density for gravity to dominate and drive the 
eventual collapse of the clump to form clusters of stars. 
We demonstrated this in the purely hydrodynamic case without magnetic fields in our most recent work
\citep{wareing19}. Therein, collapsing cold clumps formed across the resulting cloud complex, connected by
0.3 to 0.5\,pc-width filaments, with transonic flows onto the clumps and subsonic flows along the filaments.
Properties of the clumps were similar to those observed in molecular clouds, including mass, size, velocity
dispersion and power spectra. We now explore the same question in the magnetic case.
For a more complete review of research into star formation and the thermal
instability, as well the origins of this project, we refer the interested reader to two of our
previous papers from this project \citep{wareing16,wareing19}.

In Section \ref{inits}, the initial conditions are described, with a
summary of our previous work for context. In Section \ref{methods}, the 
numerical method and model are summarised. The evolution of the whole cloud is discussed in Section \ref{evol},
the formation of striations in Section \ref{strias}, the generation of
integral shaped structure in Section
\ref{ints} and hourglass-like magnetic field morphology in Section \ref{hourglasses}. The formation
of collapsing cores is presented in Section \ref{coll} and their density and velocity power spectra, as 
well as that of the whole cloud, in Section \ref{spec}. The work is summarised in 
Section \ref{conclusions}.

\section{Initial conditions}\label{inits}

In a series of recent papers we have studied the thermal instability and demonstrated the way thermally unstable
medium, under the influence of gravity and realistic magnetic fields, can evolve into thermally
stable clouds, containing warm diffuse medium and cold, high density clumps
\citep[hereafter Papers I-V respectively]{wareing16,wareing17a,wareing17b,wareing18,wareing19}.
We have used a similar initial condition across all of these works. Specifically, a diffuse cloud 
with an average number density n = 1.1 cm$^{-3}$  was
placed at the centre of the domain (0,0,0). It should be noted that we assume
a mean particle mass of $2.0\times10^{-24}$\,g (ISM 
composition of 75\% hydrogen, 25\% helium by mass) and therefore all mentions of $n$ in this work
are based on this mean particle mass. This is the same for previous works, where $n_{\rm H}$ should
now be read as $n$ defined in this manner. The radius of the cloud was typically set at 
50\,pc (e.g. Paper I) although a larger cloud radius of 100\,pc has also been investigated
(Papers IV and V). Computational domains were typically 50\% larger than the cloud.
The cloud was seeded with random density variations of 10\% about 
its average density, leading to cloud masses of 17,000\,M$_{\odot}$ and
135,000\,M$_{\odot}$. Initial pressure was set according to the (unstable) equilibrium 
of heating and cooling at P$_{eq}$/k = $4700\pm300$ K\,cm$^{-3}$, resulting
in an initial cloud temperature T$_{eq}$ = $4300\pm700$\,K. Previously, the external 
density was reduced by a factor of 10 to n = 0.1 cm$^{-3}$, but the external 
pressure matched the cloud (P$_{eq}$/k = $4700$ K\,cm$^{-3}$). The external 
medium was prevented from cooling or heating, keeping this pressure throughout 
the simulation. No velocity structure has ever been introduced to the initial condition.
When the simulation was evolved, condensations due to the thermal instability 
began to grow in the cloud and after 16\,Myrs their densities were $\sim40$ 
per cent greater than the initial average density of the cloud.
In Paper V, the hydrodynamic limit of this
model without magnetic field was explored at high resolution, showing the formation
of realistic filaments connecting a network of collapsing clumps and cores.

\begin{figure}
\centering
\includegraphics[width=85mm]{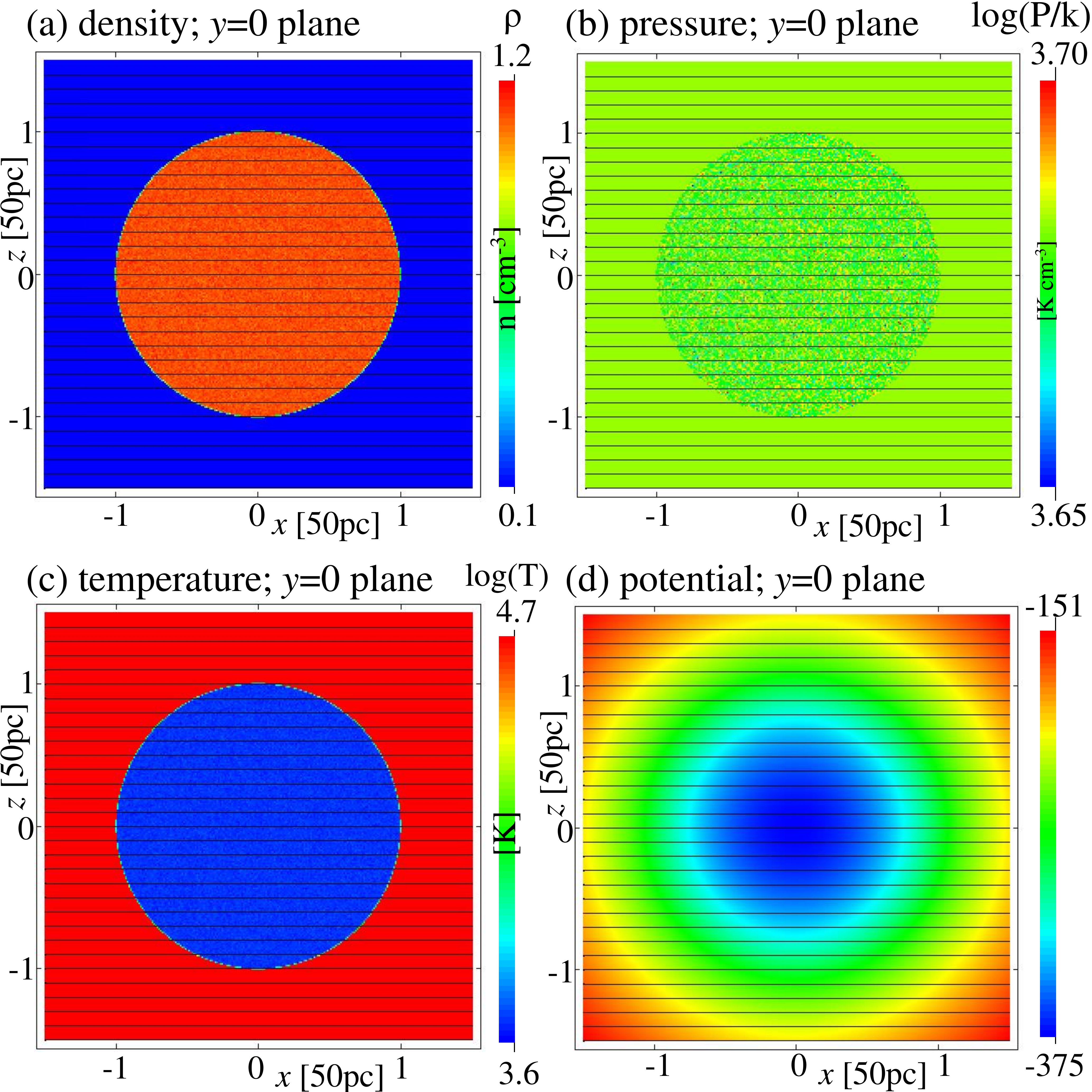}
\caption{Initial conditions in the Model 1 simulation.
Raw data are available from https://doi.org/10.5518/897.}
\label{initial}
\end{figure}

In this work, the magnetic field is included, and the 
question of whether thermally unstable material can ever lead to star formation
under only the influence of magnetic field and self-gravity
 explored for an extended range
of cloud sizes (and hence masses), external medium conditions and magnetic field
strengths. The range of conditions explored in this project are detailed in Table
\ref{table-sims}. Several individual models have been reported on previously, as noted 
in the final column of the table. Five models have been selected for this work, exploring
whether gravitationally-collapsing clumps and cores ever form.
A super-massive cloud is now included with a radius of 200\,pc and a mass of 
$1.1\times10^6$\,M$_{\odot}$. A higher density external medium has also been considered, 
specifically n = 0.85 cm$^{-3}$. This relaxes the requirement for an over-equilibrium-pressure
surrounding medium, as this density is in pressure equilibrium with the cloud, but on the warm, stable
region of the equilibrium curve. The external medium is thus allowed to heat and cool. 
Several magnetic field strengths are also explored, corresponding to plasma
beta values of 0.1, 1.0, 10, 33 and 40 - these physically correspond to $B_0$ = 3.63, 1.15, 0.363, 0.2 and 0.181$\mu$G. The
plasma beta ($\beta_{\rm plasma}$) value indicates the ratio of thermal pressure to magnetic pressure, thus these 
values can be considered as initially magnetically dominated ($<$1.0), equipartition (1.0) and 
thermally dominated ($>$1.0). Various domain sizes and
resolutions are employed in order to elucidate the interplay of magnetic field and thermal
instability. Further to the list of production simulations in Table \ref{table-sims}, extensive resolution, 
physical parameter and process tests have been performed during the project to ensure
correct capture of the effects in question.

A slice through the centre of the domain in Model 1 is shown in Fig. \ref{initial}. Model 3 is
almost identical in appearance, except that the domain and cloud radius are four times as large
and the external medium
density has the second value of 0.85. An amended scaling ratio in the code is used to allow the
cloud to extend only to 1 in code units, but a higher resolution employed to allow comparison 
with earlier Models. Model 5 concerns an extracted central section from Model 3, initialised
in a very similar manner to the study reported in Paper V. 
A central spherical region of Model 3, radius 50\,pc, was placed in a cube 
domain 120\,pc on a side. The stationary surroundings outside the spherical region were pressure 
and density matched to those low-density surroundings inside the region in the same manner as 
Paper V. A uniform magnetic field, equivalent to the field strength in and around the sheet
measured from Model 3, was inserted across the domain. Model 5, with initially 5 levels of adaptive mesh refinement (AMR),
matched the finest levels of resolution with Model 3, but as Model 5 was evolved, extra levels of AMR
were rapidly added to accurately resolve the formation of massive clumps, resulting in the finest resolution as described
in Table \ref{table-sims}.

\section{Numerical methods and model}\label{methods}

The MG MHD code has been used here in the same manner with the
same heating and cooling prescriptions as throughout 
Papers I to V and for reasons of brevity, we ask the interested
reader to consult those papers for further details.

\begin{table*}
\begin{center}
  \caption{Details of the suite of 3D Cartesian MHD simulations, including heating, 
cooling and self-gravity, performed during the program of work reported in Papers I-V and herein. 
Individual simulations discussed in this work are given specific Model names. Specific citations are 
given to simulations discussed in preceding publications.}
  \label{table-sims}
  \begin{tabular*}{176mm}{lllllllllll}
\hline
Name & Physical & Cloud & Cloud & Initial & $\rho_{\rm{surr.}}$ & G0 & Max. & Finest & References/Comments\\
herein& Domain & Radius & Mass & $\beta_{\rm{plasma}}$ & n & & levels & Res. \\
          & [pc on a side]& $R_{init}$ [pc] & [M$_{\odot}$] & & [cm$^{-3}$] & [\# of cells]& AMR & [pc] \\
\hline
             & 150 & 50   & 1.7e4 & 0.1   & 0.1 & $4\times4\times4$ & 8 & 0.29 & See Paper I\\ 
Model 1 & 150 & 50   & 1.7e4 & 1.0   & 0.1 & $4\times4\times4$ & 8 & 0.29 & See Papers I and II\\ 
             & 150 & 50   & 1.7e4 & 10.0 & 0.1 & $4\times4\times4$ & 8 & 0.29 & \\
             & 150 & 50   & 1.7e4 & 1.0   & 0.85 & $4\times4\times4$ & 8 & 0.29 & \\ 
             & 150 & 50   & 1.7e4 & 10.0 & 0.85 & $4\times4\times4$ & 8 & 0.29 & \\ 
             & 150 & 50   & 1.7e4 & 33.0 & 0.85 & $4\times4\times4$ & 8 & 0.29 & \\ 
\hline
Model 2 & 300 & 100 & 1.35e5 & 1.0   & 0.1 & $8\times8\times8$ & 8 & 0.29 & See Paper IV\\ 
             & 300 & 100 & 1.35e5 & 10.0 & 0.1 & $8\times8\times8$ & 8 & 0.29 & \\
             & 300 & 100 & 1.35e5 & 1.0   & 0.8 & $8\times8\times8$ & 8 & 0.29 & \\ 
             & 300 & 100 & 1.35e5 & 10.0 & 0.8 & $8\times8\times8$ & 9 & 0.14 & \\
\hline
             & 600 & 200 & 1.1e6  & 0.1 & 0.85 & $6\times6\times6$ & 8 & 0.78 & \\
Model 3 & 600 & 200 & 1.1e6  & 1.0 & 0.85 & $6\times6\times6$ & 9 & 0.39 & \\ 
Model 4 & 600 & 200 & 1.1e6  & 10.0 & 0.85 & $6\times6\times6$ & 9 & 0.39 & \\ 
             & 600 & 200 & 1.1e6  & 40.0 & 0.85 & $6\times6\times6$ & 8 & 0.78 & \\ 
\hline
Model 5 & 120 & 50 & 1.38e5 & $\sim1.0,$ & 0.385 & $12\times12\times12$ & 8 & 0.078 & Extracted from Model 3\\ 
& & (disc) & & up to 2.8& & & & & after collapse to a disc \\
& & & & at high $\rho$ & & & & & $t_{\rm start}$=50.1\,Myrs \\
\hline
\end{tabular*}
\end{center}
\end{table*}

Table \ref{table-sims}  presents details of the  simulations presented
in this work. The simulations employ  a varying base AMR grid (G0) and
varying  levels  of AMR.   The  finest  grid resolution  is  typically
0.29\,pc,  increased to  0.078\,pc  for the  high-resolution study  of
collapsing cores to be comparable to Paper V. G0 needs to be coarse to
ensure fast  convergence of  the MG Poisson  solver.  
It should be  noted that here, as  in our previous papers,  we do not
  include thermal conductivity, nor do  we resolve the Field length in
  the manner discussed by \cite{koyama04}. As we show in a convergence
  study and discussion in Appendix  \ref{sectapp1}, it is not strictly
  necessary to  do either of  these things,  if we are  only concerned
  with the large-scale behaviour of  the thermal instability, which is
  convergent  at  such  scales.   Any  small-scale  structure  introduced by  not
  including  thermal   conductivity  is   subsumed  into   the  larger
  structure.

The central region of the Model 3 simulation was extracted in order to provide the initial condition
for resimulation in Model 5. Mapping of the region extracted from Model 3 onto Model 5 was 
performed in a simple linear fashion over all three coordinate directions for every cell in question. 
The dense sheet structure in this grid was well-resolved (by 10 or more cells) in order to ensure 
no loss of detail. Model 5 started with 5 levels of AMR, but extra levels were added as the structure
reached resolution limits for each level, up to 8 levels of AMR. The simulations were typically 
run across 96 cores of the ARC HPC facilities at the University of Leeds, including some of 
the final available cycles of the long-lived and well-used DiRAC1 UKMHD machine that has been
particularly beneficial to this project (see also the Acknowledgments). 

\begin{figure*}
\centering
\includegraphics[width=150mm]{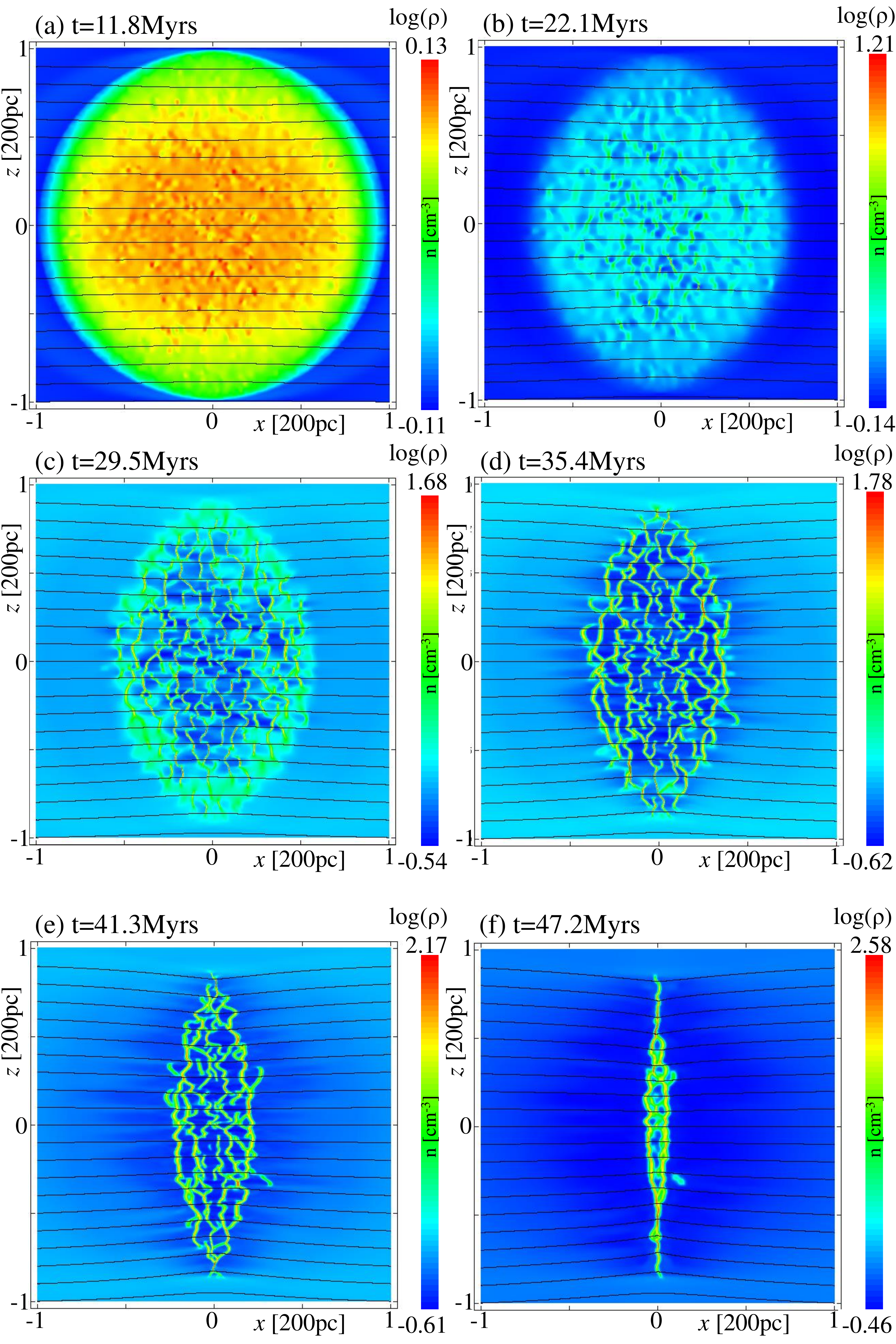}
\caption{The evolution of Model 3. Number density slices
    through the $y=0$ plane are shown with magnetic field lines.
    Raw data are available from https://doi.org/10.5518/897.}
\label{evolution}
\end{figure*}

\section{Results and Discussion}\label{results}

Cloud structure and evolution are discussed in this section. In particular, time-variation 
of properties, derived statistics, slices and column density projections at snapshots
in time are examined. MG plotting mechanisms and in-house software 
are used to visualise these data.

\subsection{Evolution}\label{evol}

The evolution of the Model 1 cloud is discussed in detail in Paper I, where figs. 4, 5, 6 and 7
illustrate the evolution. This evolution can be summarised as follows.
From the unstable initial condition, as is shown for Model 3 in Fig. \ref{evolution},
over-dense thermally stable, cold condensations begin to appear after approximately 20\,Myrs (Fig. \ref{evolution}b), alongside
regions of diffuse, stable warm medium. Thermal motions of the gas along the
field-lines accompany this phase. The collapse of the cloud under gravity leads to greater sub-Alfv\'enic 
motions along the field-lines toward the minimum of gravitational potential along that field line, in
agreement with the theoretical predictions and simulations \citep{giri18}. Magnetic pressure 
supports the cloud across the field lines. Condensations appear to grow across the field-lines 
(Fig. \ref{evolution}c), but this is the result of material collecting through motion along the field-lines, not
motion across the field-lines. This has the effect of increasing density in some locations and 
creating low-density field-aligned structure in others, resembling striations inside and outside the cloud (Fig. \ref{evolution}d). Note
the consistent appearance of such structure in the convergence tests shown in Appendix \ref{sectapp1}.
The cloud resembles a transitory foam-like structure (Fig. \ref{evolution}e) on the way to collapsing to a thick,
corrugated, approximately circular sheet after 40\,Myrs, with a radius the same as the initial 
condition (Fig. \ref{evolution}f). In projection, the corrugations across the sheet can appear to give an integral-shaped 
morphology, as demonstrated by Model 2 and discussed later. The sheet radius is 50\,pc in the 
case of Model 1 and 100\,pc in the case of Model 2. If enough mass is present in the sheet 
for gravity to overcome the magnetic pressure support, the cloud then globally collapses
across the field lines at the same time as approaching a thick sheet-like configuration. Model 1
does not collapse across the field lines, whereas Models 2, 3 and 4 do to lesser (Model 3) and greater (Models 2 and 4) effect. This
process intensifies the magnetic field, resulting in hourglass field morphology (discussed
later in more detail for Model 4). If enough mass is present locally (originally in the column along
the magnetic field line), then magnetic supercriticality can be achieved locally and gravity overcomes
the magnetic support leading to gravitational collapse of the clumps into cores on a shorter free-fall 
time-scale than that of the whole cloud. This is demonstrated in Model 3 and at a higher resolution by Model 5.  

\begin{figure}
\centering
\includegraphics[width=85mm]{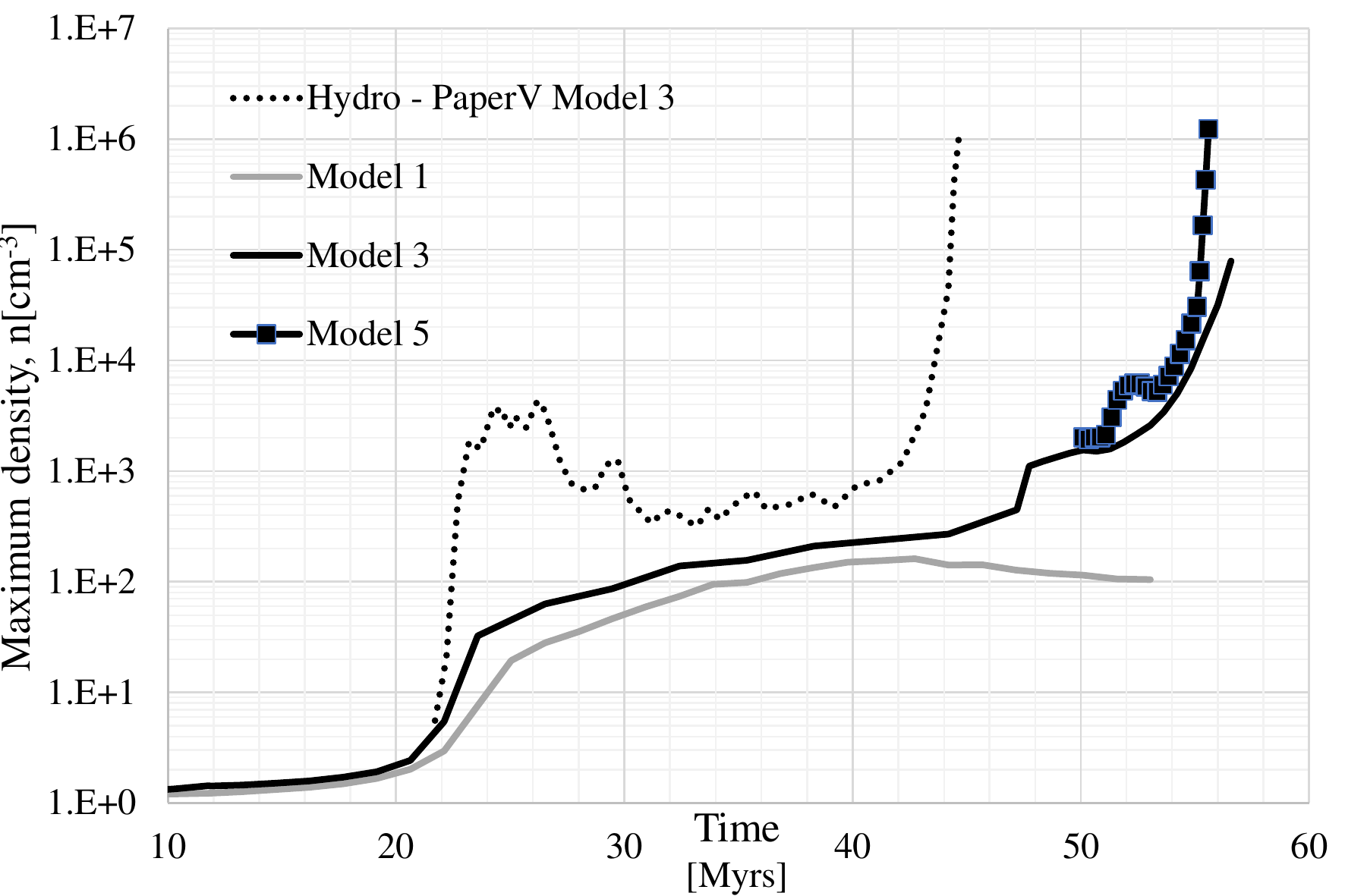}
\caption{The evolution of the maximum density. Presented are the collapsing HD case from Paper V
and a number of the MHD cases: the low mass magnetic cloud (Model 1), the high-mass cloud (Model 3) 
and the resimulation of an extract of the high-mass cloud at higher resolution (Model 5). 
Raw data are available from https://doi.org/10.5518/897.}
\label{maxrho}
\end{figure}

Fig. \ref{maxrho} shows the evolution of the maximum density in the simulations - a simple way to
track if the cloud creates clumps and cores, or remains quiescent. For reference, Fig. \ref{maxrho} shows the
behaviour of the purely HD case that creates clumps and cores in the case of no magnetic
field, detailed in Paper V. In this HD case, there is a phase of evolution where there are reasonably high
peaks in the maximum density (23-30\,Myrs), but as noted in detail in Paper V, the dynamics of the cloud overcome 
any possibility of forming truly bound and collapsing clumps until later. Characteristic of this phase is a network of clumps
connected by filaments. Both clumps and filaments display properties that compare well to observations,
specifically filament widths, temperatures and flow characteristics and clump size-scales,
masses, temperatures and velocity dispersions. For the details of these comparisons, please refer to Paper V.
Turning now to the MHD cases, it is first reassuring to note that the HD and MHD models are converged
until approximately 22\,Myrs, capturing the initial evolution of the thermal instability. 
The HD and MHD cases then diverge. There is no phase of dynamic evolution in either of the low mass (Model 1) or high 
mass (Model 3) MHD cases. Any formation of a filamentary network is suppressed. The thermal flow that 
condensed the thermally stable structure in the HD case is now confined to 1D along the field lines by the magnetic pressure.
Both magnetic cases follow a similar evolution until approximately 40\,Myrs. Model 1 does not have 
sufficient mass to either globally collapse across the field lines under the effect of gravity, or for the 
formation of individual clumps across the sheet. However, Model 3, with 64 times more mass than
Model 1 (see Table \ref{table-sims}) clearly does have enough mass to overcome the magnetic support
and collapse. The collapse of the cloud continues with the same power-law index (constant gradient in the plot), in terms of maximum density increase
with time, until with enhanced resolution, Model 5 reveals the collapse of individual clumps to high density.
The power-law indices (gradients) of the HD and MHD cases are very similar: compare the period of the HD case
between 34 and 40\,Myrs to that of the MHD case between 30 and 45\,Myrs. Note also that both require
regions with density over 1000\,cm$^{-3}$ before clumps can collapse. Above this density, 
both HD and MHD cases display similar clump collapse time-scales in this figure, albeit occurring at 
54-56\,Myrs in the MHD case as compared to 42-44\,Myrs in the HD case. Clearly in the HD case, the nature
of the cloud evolution unconstrained by the magnetic field leads to flows in any direction and collapse to
higher density on a shorter timescale, even if there is a period where the thermally-induced flows prevent gravitational collapse.
Gravity now dominates the final evolution in both the HD and high-mass MHD cases. 

To understand why Model 3 collapses and Model 1 does not, we now explore the critical differences 
between the models. The initial cloud is in an unstable equilibrium state. On the timescale of 
approximately 20\,Myrs (Fig. \ref{evolution}b) 
it evolves  into a thermal  equilibrium state containing  dense,  cold  stable  material and  
diffuse,  warm  stable material in pressure equilibrium, as also illustrated in
fig. 2 of Paper V. The final evolutionary outcome can be 
illuminated by considering the equilibrium state produced by the collapse of a uniform, non-rotating,
isothermal, spherical core as considered by \cite{mousch76a,mousch76b} and later in terms of
MHD shocks and subcritical cores relevant to this work by \cite{vaidya13}. Specifically, for a 
zero-temperature core, \cite{mousch76c} derive the critical value of mass-to-flux ratio above which 
a core will collapse under gravity as
\begin{equation}
\frac{{{M_{\rm crit}}}}{{{\Phi _{\rm crit}}}} = \frac{{0.53}}{3\pi }{\left( {\frac{5}{G}} \right)^{1/2}}.
\end{equation}
\cite{vaidya13} note that for the case of ideal MHD the mass-to-flux ratio does not change and this 
becomes
\begin{equation}
\frac{M}{\Phi } = \frac{{4{\rho_i}{R_i}}}{{3{B_0}}}.
\end{equation}
Thus, in this case with a constant density for all initial conditions within the thermally unstable range,
there is a critical initial cloud radius, below which the cloud will not collapse, given by
\begin{equation}
{R_{crit}} = \frac{{3{B_0}}}{{4{\rho_i}}}\frac{M_{\rm crit}}{\Phi_{\rm crit}}.
\end{equation}
For $\rho_i=1.1$\,cm$^{-3}$ and $\beta_{\rm plasma}=1.0$ resulting in B$_0$=1.15\,$\mu$G, we obtain
R$_{\rm crit}$=73\,pc. Model 1, with a radius of 50\,pc, is therefore sub-critical. Models 2, 3 \& 4
are supercritical and will eventually collapse.

\begin{figure*}
\centering
\includegraphics[width=160mm]{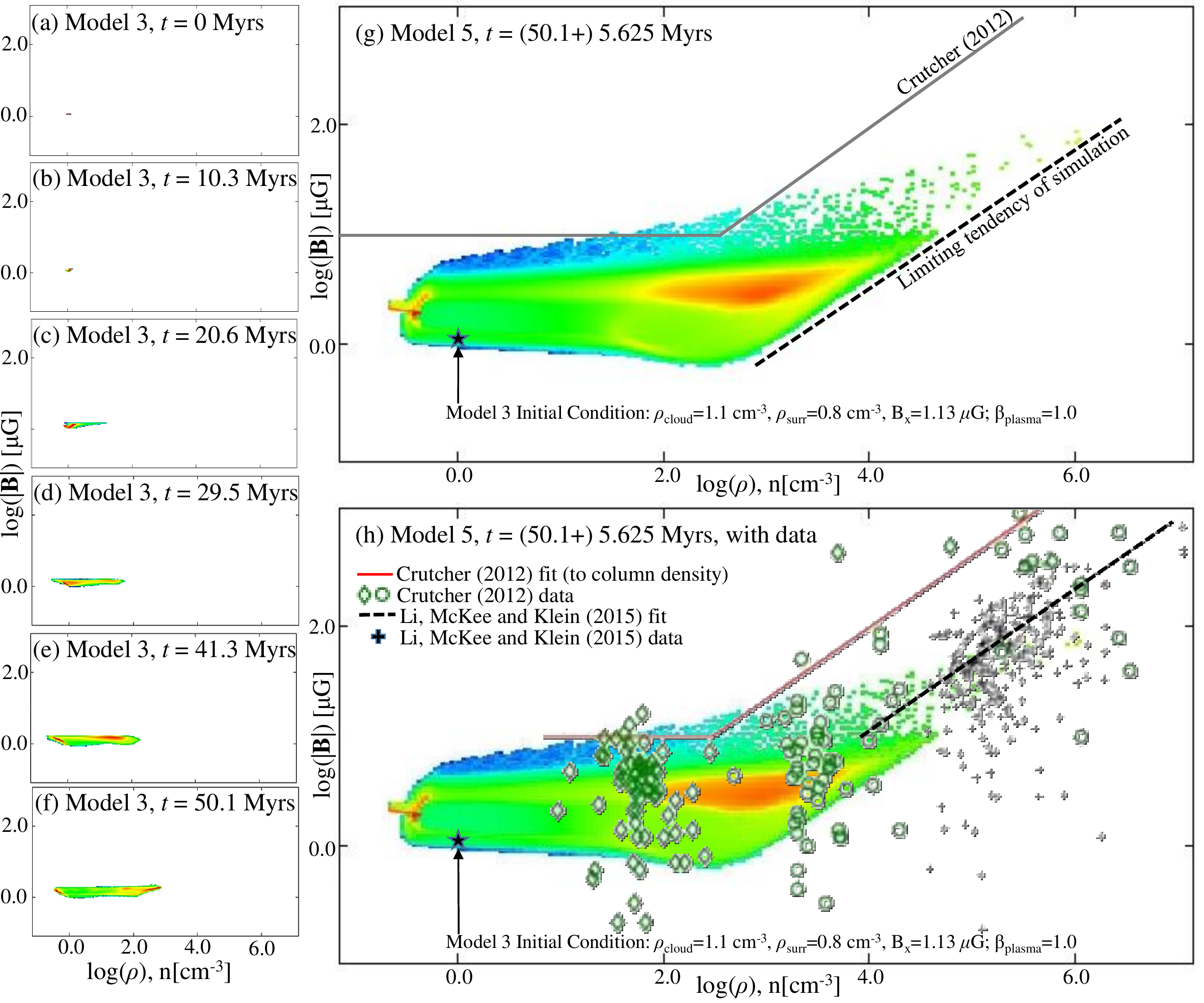}
\caption{Comparison of Models 3 and 5 to observational data, fits and another simulation concerning
the relationship between magnetic field strength and data. Panels (a) to (f) show the evolution of
the distribution for Model 3. Panel (g) shows a comparison between Model 5 and the Bayesian fit 
for maximum field strength presented by \citet{crutcher10} and \citet{crutcher12}. The dashed line shows the 
indicative limiting power-law trend of the minimum field strength for a certain density in the simulation.
Panel (h) shows the same Model 5 distribution as in (g), but overlaid with the Bayesian fit and data used
to obtain the fit \citep{crutcher12}, as well as simulated data and a fit from the turbulent simulations of 
\citet{li15}. Raw data are available from https://doi.org/10.5518/897.}
\label{crutcher}
\end{figure*}

\cite{crutcher12} showed (in their fig. 6) the application of a Bayesian statistical technique
to analyze samples of clouds with Zeeman observations \citep{crutcher10}. The model assumed that the
maximum magnitude of the magnetic field was independent of density up to a certain value of density.
The Bayesian analysis fit to observational data from four surveys for the Zeeman, H{\sc i}, OH 
and CN data revealed the maximum magnetic field strength to be approximately B$_0 \sim$ 10$\mu$G up to approximately
n$_0 \sim$ 300\,cm$^{-3}$ (n(H$_2) \sim 150$\,cm$^{-3}$). Above this, maximum field strength was assumed to
have a power-law increase, which was found to be of the form B $\propto$ B$_0$(n/n$_0$)$^{0.65}$, indicative
of isotropic contraction and magnetic supercriticality. 
More recently, \cite{tritsis15} found a power-law index of $\sim 0.5$ indicative of 
one-dimensional collapse along fieldlines based on an analysis of a limited
set of Zeeman data and \cite{kandori18} found an index of $0.78 \pm 0.10$ toward the starless dense 
core FeSt 1-457.

\begin{figure*}
\centering
\includegraphics[width=160mm]{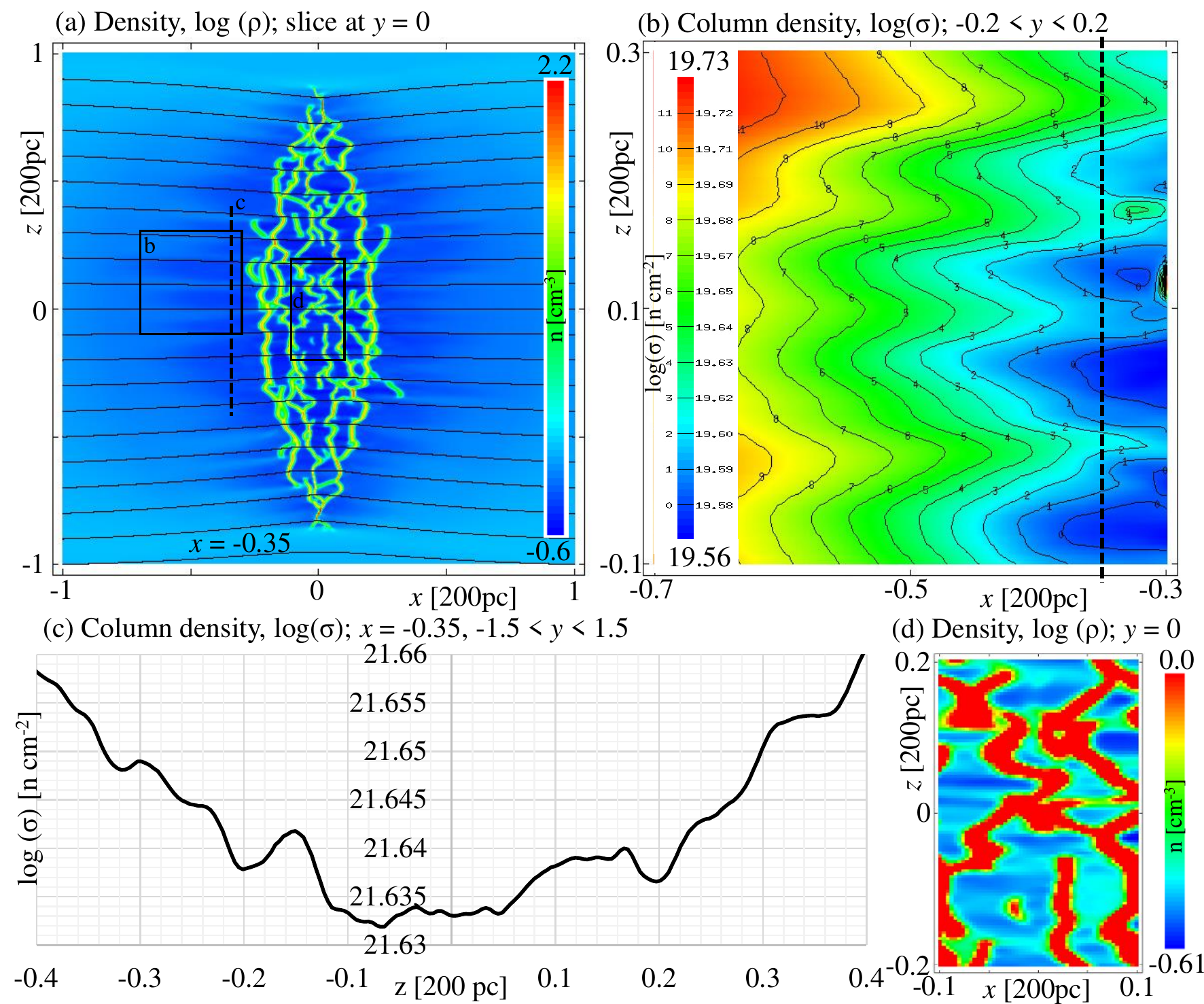}
\caption{The appearance and properties of striations closely-aligned to the magnetic field in Model 3
after 41.3 Myrs of simulated evolution. Panels (b), (c) and (d) show the column density for the regions marked 
'b' and 'd' and the line marked 'c'  in panel (a).
Raw data are available from https://doi.org/10.5518/897.}
\label{striations}
\end{figure*}

In Fig. \ref{crutcher}, we now explore the behaviour of the simulations with respect to the \cite{crutcher12}
relationship. Fig. \ref{crutcher} shows the magnetic field strength versus density distribution for Model
3 and its high resolution resimulation in Model 5. In the colour distribution plots, every cell in the grid 
simulation has been accounted for and the colour scale indicates the frequency of cells with such magnetic 
field and density properties, from blue indicating very few cells to red indicating a large number of cells. 
Given that the initial magnetic field strength is constant and the initial density only varies over a narrow
range, the distribution of the initial condition as shown in Fig. \ref{crutcher}a is very confined. Over
the next 50\,Myrs as shown in panels a to f (the entire duration of Model 3) the major effect on the 
distribution is a rightwards spread - a range in density of more than 3 orders of magnitude, but very 
narrow range of slightly increased magnetic field strength.
As can be seen in Fig. \ref{crutcher}f, increasing density above a certain level is then also
accompanied by increasing magnetic field. It is interesting to note that this density threshold (300\,cm$^{-3}$)
 is the same as that at which \cite{crutcher12} deduced a switch 
from a flat maximum magnetic field strength to a power-law relationship between maximum magnetic field 
strength and density. Only when gravitationally unstable regions appear and collapse to high densities 
does the magnetic field strength increase with density, as can be seen in Fig. \ref{crutcher}g. The 
simulations appear to show the same tendency to a power-law increase with the same gradient index.
The distribution falls reassuringly below Crutcher's maximum field strength relationship.
It should be noted that different initial conditions - specifically an initial magnetic field greater than 10\,$\mu$G 
in the thermally unstable medium - would result in a distribution above Crutcher's relationship at
densities below 10$^3$.

It is notable, as shown in Fig. \ref{crutcher}h, that when the observational
data are included (green circles and diamonds), the spread of the data is wide and encompasses the 
simulated distribution. The peak of the simulated distribution is in good agreement with the majority
of the observational data. The black crosses fit by the dashed line in Fig. \ref{crutcher}h are
the simulated data and fit of \cite{li15}, who consider the turbulent formation of molecular clouds. They
conclude the results of their strong field model (Alfv\'en Mach number of 1) are in very good quantitative
agreement with observations. On this point in particular, their work is in good agreement with our simulated distribution at high
density, even though they start from a very different premise of turbulence-driven molecular cloud
formation. \cite{crutcher12} derives the relationship as the limiting relationship between magnetic
field and density and it would appear that this is indeed the case.

Everything together  in Fig. \ref{crutcher}h  - data, fit to  data and
two  simulations from  very different  initial conditions  - seems  to
agree well  with the  conclusions of  \cite{crutcher12}: specifically,
that magnetic  field morphologies  appear to  be generally  smooth and
coherent  across the range  of scales considered here
  from  10$^2$ to  10$^{-2}$\,pc.  However  it should  be noted  that we
  impose a  smooth initial magnetic  field in our  initial condition.
Ordered   fields  perpendicular   to   elongated  structures   suggest
contraction along  magnetic field lines. Data,  both observational and
simulated, seems generally consistent with the scenario that molecular
cloud complexes  are formed by  accumulation of matter  along magnetic
flux tubes, such  that the magnetic field does not  increase much with
density up  to n $\sim  300$\,cm$^{-3}$. Above  this density,
\cite{crutcher12}  concludes, magnetic  field strengths  increase with
increasing density, with a  power-law exponent indicating that gravity
dominates magnetic  pressure. Our simulations indeed  become dominated
by gravity above this density in precisely this manner, providing some
clarity of the astrophysical processes at work which can reproduce the
observations.

\begin{figure*}
\centering
\includegraphics[width=160mm]{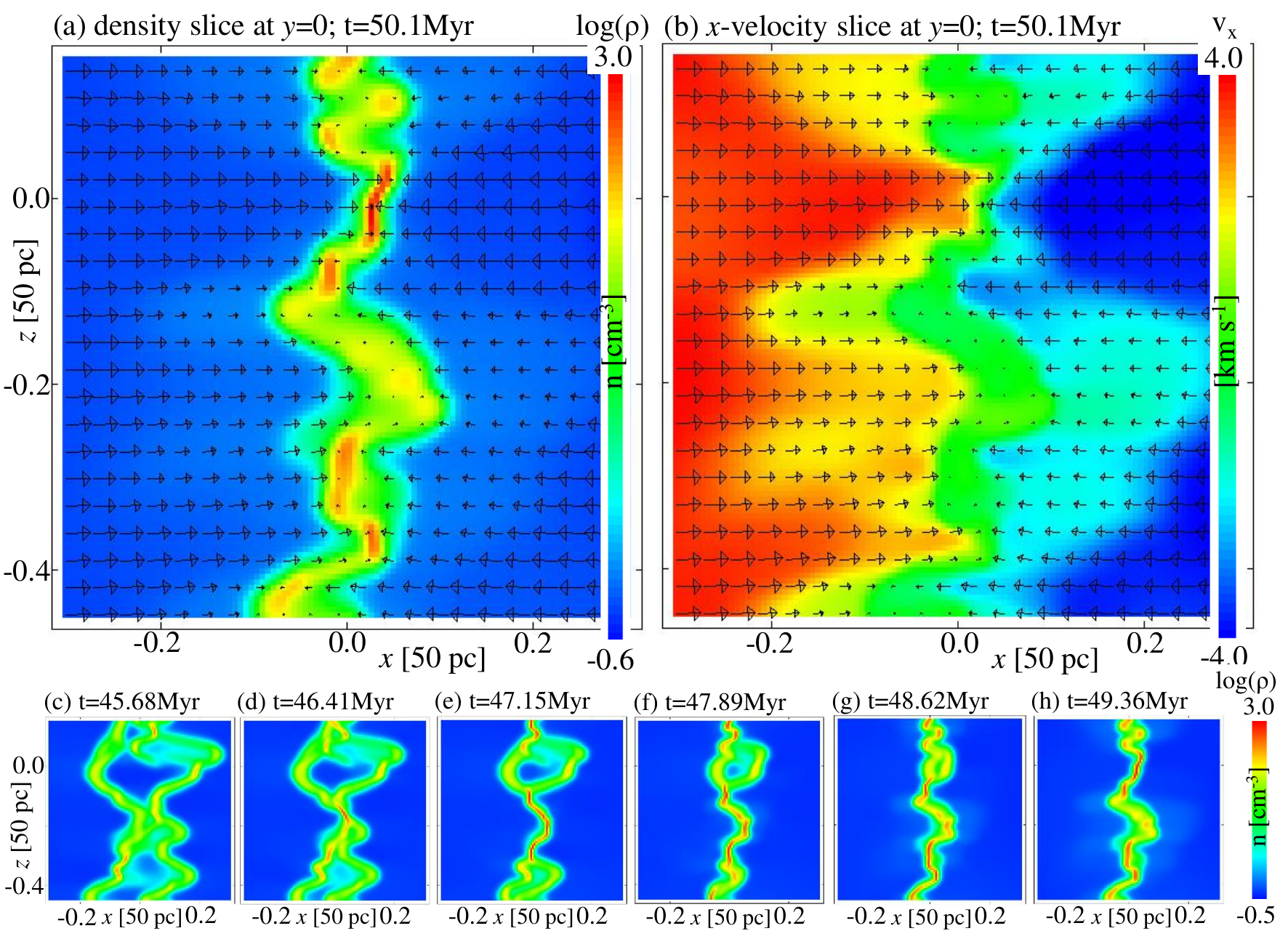}
\caption{The appearance of integral-shaped structure and the associated velocity field
in Model 2. Reversals in velocity vector over short 
distances close to the dense structure are apparent in panel (a), whilst the colour map
of panel (b) demonstrates the red/blue-shift effect on either side of the dense structure.
Panels (c) to (h) illustrate the timescale and motion of collapse.
Raw data are available from https://doi.org/10.5518/897.}
\label{integrals}
\end{figure*}

\subsection{Striations}\label{strias}

In Fig. \ref{striations}, an example of the striations that form around the dense cloud structure 
is presented. Fig. \ref{striations}a shows a slice through the centre of the simulation domain of
Model 3, where striations - low-density, field-aligned structure - can be seen outside the dense
regions of the cloud. These striations are in the low-density warm diffuse medium: their density 
is below n $\sim 1$\,cm$^{-3}$. In these simulations, at this resolution, they do not
contain any cold, dense material. Figs \ref{striations}b and \ref{striations}c show the typical
column density variation across these striations. The average difference peak to trough in 
column density is around 10-15\%, but this does range up to 25\% just in the panel shown here.
Note that the column density in Fig. \ref{striations}c is larger than that in Fig. \ref{striations}b 
as this cut takes account of the depth of the entire simulation box (3 in code units), so the 
variations peak to trough are smaller in terms of percentage difference given as they appear as 
a variation on this background created by the whole domain, but the column density in panel c 
is in agreement with the range observed around n$_{\rm H} \sim 10^{21}$\,cm$^{-2}$ for structure in
Taurus at a distance of $\sim$120\,pc \citep{goldsmith08,tritsis16}. 

It should also be noted that low-density striations are also found in the centre of the Model 
3 simulated cloud, as shown in Fig. \ref{striations}d, and across the entire range of magnetic 
simulations performed during this project (see Table \ref{table-sims}, Papers I, II and IV). Such 
striations are essentially the same as those outside the cloud, though they
are not physically connected. It is clear from Figs \ref{striations}a and \ref{striations}d 
that the low-density striations are associated with structure in the dense filaments e.g. local 
high densities or intersections of filaments. We also find that the striations only appear after  
that of dense structure, which suggests that they are generated by the thermally-induced flows that cause the  
dense structures; compare panels (c) and (d) of Fig.\ref{evolution} and note
the material trailing from the accretion that remains collimated along the field-lines. This
interpretation is supported by the velocity structure of the striations, indicating the remnant
of the accretion flow. This could be due to the same mechanism of MHD waves discussed by
\cite{tritsis16} for high density striations, as the comparatively long-lived low-density striations in the
magnetic simulations are clearly magnetically controlled (as opposed to those that appear briefly
in the purely hydro simulations in Paper V, show no preferred direction and lead to the filaments 
that connect clumps). Further work is required to explore this possibility,
but the presence of low-density striations in simulations with and without self-gravity,
with and without periodic domains (see Appendix \ref{sectapp1} and Paper I for such periodic 
simulations), lends credibility to an origin in the thermal instability.

In general, the MHD simulations indicate a preference for low-density structure (striations) to align
parallel to the magnetic field and high-density structure (sheets) to orient perpendicular to the magnetic
field. This is in agreement with observations, such as a {\it Mopra} telescope survey of molecular
rotational lines towards the young giant molecular cloud Vela C \citep{fissel19} which quantified
the orientation of gas structure with respect to the cloud magnetic field orientation. Those authors
estimate the characteristic transition density from parallel to perpendicular to be around 
$10^3 \pm 10^1$\,cm$^{-3}$, similar to what we observe in our simulations. However, we must
stress that the striations we observe in our simulations are not the same as those dense, cold striations observed in CO
emission and modelled by \cite{tritsis16}. Our diffuse, warm striations are more likely to be
linked to the elongated fiber structures observed at high Galactic latitudes in the diffuse interstellar
medium, referred to by \cite{tritsis16}.

It is possible that AMR can produce striation-like grid-aligned structure, if a faulty grid control algorithm 
is present and careful consideration of the results is not employed. We have
seen this in other applications outside astrophysics and responded by reconsidering the grid control
algorithm and repeating the calculations. No such amendments have been necessary in the suite of 
simulations presented in this series of papers. The striations are well-resolved  
perpendicular to their orientation by 5 to 10 cells. Their initial appearance in the simulations
follows the appearance of thermally-stable structure, not any changes in grid resolution in that local
area. Their appearance is also common at the same size-scale across our results presented herein, 
the convergence tests presented in Appendix \ref{sectapp1} with and without thermal conductance 
and across the suite of simulations in Papers I, II and IV as already noted.

\begin{figure*}
\centering
\includegraphics[width=160mm]{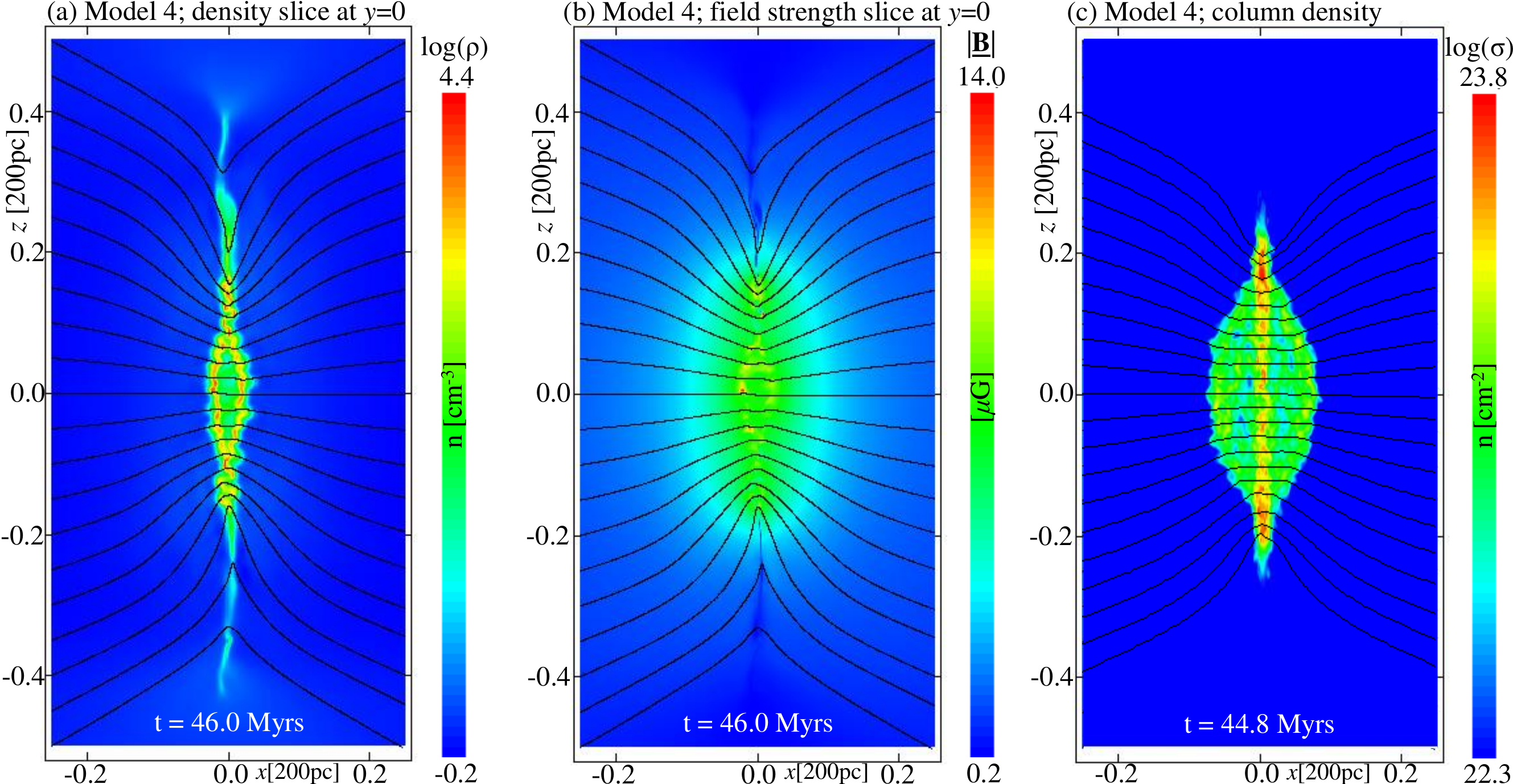}
\caption{Demonstration of global cloud collapse and formation of hourglass-like field structure in
Model 4. The initial simulation had a uniform magnetic 
plasma beta of 10 across the domain. At the times presented, the plasma beta now varies
across the simulation volume and realistic values of field magnitude (5-10 $\mu$G) and plasma beta ($\sim1$)
are obtained in the high density cloud regions, regardless of the initial condition. In projection (c),
two approaching `clumps' appear to be dragging the field akin to recent observations in OMC1 
(Pattle et al. 2017). Raw data are available from https://doi.org/10.5518/897.}
\label{hourglass}
\end{figure*}

\subsection{Integrals}\label{ints}

In the simulations, the thick disc-like molecular clouds have a natural corrugation which appears 
as an integral-shape in cross-section and projection. The disc briefly oscillates about the potential minimum along the
field lines, with a standing wave nature perpendicular to the magnetic field
and a peak-to-peak distance set by the size-scale of
5-10\,pc that results from the thermal instability \citep{falle20}. Fig. \ref{integrals} shows this effect 
in the Model 2 simulation once the cloud has collapsed to a sheet but before the disc 
reaches static equilibrium at the minimum of the gravitational potential. This is a 
transient effect of `collapse-overshoot' as the material flows non-uniformly along the fieldlines into the gravitational
potential well as a result of the thermal instability. The velocity range is not insignificant, from -4 to +4\,km\,s$^{-1}$ 
as shown in Fig. \ref{integrals}b. These structures and the disconnects in velocity are similar to
those presented in \cite{stutz18} and \cite{lobos19} concerning a wave-like model for the ISF in Orion A.
Clearly there is a realistic physical
formation mechanism for such structure and velocity patterns, deriving from the thermal instability
and leading to interaction between magnetic field and gravity. Similarly, it is
therefore possible that the `vibration' of Musca detected by \cite{tritsis18} is an effect of 
gravitational contraction and oscillation about the potential minimum over a period of Myrs
(panels (c) to (h) of Fig. \ref{integrals}),
with striations formed in the manner discussed above. Further work to investigate this is required.

Interestingly, \cite{liu19} investigate the internal gas kinematics of the filamentary cloud
G350.54+0.69 and find a large-scale periodic velocity oscillation along the filament, with a
wavelength of 1.3\,pc and an amplitude of $\sim$0.12\,km\,s$^{-1}$. The authors 
conjecture the periodic velocity oscillation could be driven by a combination of longitudinal
gravitational instability and a large-scale periodic physical oscillation along the filament,
possibly an example of an MHD transverse wave. It's clear that during the collapse
of a thermal instability driven cloud, the resulting velocities encompass this 
observation. In our simulations, the velocity range reduces over time as the amplitudes
of the oscillation reduce due to the sheet reaching the potential minimum. Our model
and the observations of \cite{liu19} are therefore compatible. 

In addition, \cite{shi19} concluded from red-shifted and blue-shifted velocity patterns around the 
Taurus B211/B213 filament that the filament was 
initially formed by large-scale compression of H{\sc i} gas and is now growing in mass by 
gravitationally accreting molecular gas from the ambient cloud. The simulations show similar 
global red-shifting and blue-shifting patterns on either side of the molecular structure, as
can be inferred from the velocity patterns in Fig. \ref{integrals}b. The observed shifted
velocities around the Taurus B211/B213 filament are on the order of 1-2\,km\,s$^{-1}$ -
well within the range of velocities that we see in our simulations.

In summary, the gravitational collapse of a cloud triggered by thermal instability produces 
a cloud with a converging flow which is similar to models involving colliding flows. However,
they do not represent large-scale converging flows, but instead are a natural consequence  
of the thermal instability combined with gravity. This work now
adds a further possibility of thermal flow origins to such observed velocity patterns and 
structures.

\begin{table*}
\begin{center}
  \caption{Properties of the 33 clumps identified by 
  the FellWalker algorithm, at t=55.6 (50.1 + 5.5) Myrs in the Model 5 simulation.
  Snapshots of slices through the clumps are available from https://doi.org/10.5518/897.}
  \label{tableclumps}
  \begin{tabular*}{\textwidth}{lllllllllll}
\hline
 &  M$_{\rm total}$ & M$_{\rm warm}$ & M$_{\rm unstable}$ & M$_{\rm cold}$ &  $\rho_{\rm max}$ & T$_{\rm min}$ & Scale & v$_{\rm disp}$ & Bound? & Jeans \\
 & [M$_{\odot}$] & [M$_{\odot}$] & [M$_{\odot}$] & [M$_{\odot}$] & n [cm$^{-3}$]& [K] & [pc] & [km\,s$^{-1}$] & & unstable?\\
\hline
  1 & 1.98e3 & 5.81e0 & 3.24e0 & 1.98e3 & 4.41e3 & 16.4 & 4.0 & 0.27 & N & N \\ 
  2 & 1.78e3 & 4.35e0 & 3.21e0 & 1.77e3 & 9.42e3 & 14.7 & 2.0 & 0.26 & N & N \\ 
  3 & 8.19e3 & 9.28e0 & 1.56e1 & 8.18e3 & 5.20e3 & 19.8 & 2.0 & 0.12 & Y & Y \\ 
  4 & 1.41e3 & 2.00e0 & 2.72e0 & 1.40e3 & 2.60e4 & 12.8 & 1.0 & 0.28 & Y & Y \\ 
  5 & 1.47e3 & 4.36e0 & 2.37e0 & 1.46e3 & 1.24e6 & 8.1 & 2.0 & 0.73 & Y & Y \\ 
  6 & 3.01e3 & 9.26e0 & 5.48e0 & 3.00e3 & 1.40e4 & 13.9 & 3.0 & 0.28 & Y & Y \\
  7 & 2.49e3 & 2.55e0 & 4.49e0 & 2.49e3 & 3.13e3 & 17.6 & 2.0 & 0.21 & Y & Y \\
  8 & 3.69e3 & 6.65e0 & 7.26e0 & 3.68e3 & 2.94e3 & 17.8 & 3.0 & 0.19 & Y & Y \\
  9 & 2.88e3 & 5.08e0 & 5.28e0 & 2.88e3 & 3.22e3 & 18.1 & 3.0 & 0.15 & N & N \\
10 & 1.63e3 & 2.48e0 & 3.45e0 & 1.62e3 & 2.57e3 & 18.6 & 5.0 & 0.17 & Y & Y \\
11 & 2.85e3 & 3.83e0 & 5.08e0 & 2.85e3 & 3.03e3 & 17.7 & 2.0 & 0.17 & N & N \\
12 & 1.76e3 & 2.09e0 & 3.48e0 & 1.75e3 & 7.83e3 & 15.1 & 1.5 & 0.20 & Y & Y \\ 
13 & 1.72e3 & 2.02e0 & 3.49e0 & 1.72e3 & 1.46e4 & 13.8 & 1.5 & 0.22 & Y & Y \\ 
14 & 7.46e2 & 8.15e-1 & 1.41e0 & 7.44e2 & 3.15e4 & 12.6 & 1.0 & 0.26 & Y & Y \\ 
15 & 2.36e3 & 2.07e0 & 4.73e0 & 2.35e3 & 6.25e3 & 15.7 & 1.5 & 0.23 & N & N \\
16 & 3.43e2 & 3.24e-1 & 6.84e-1 & 3.42e2 & 5.86e3 & 15.9 & 3.0 & 0.23 & N & N \\ 
17 & 3.97e3 & 5.55e0 & 7.57e0 & 3.97e3 & 7.64e3 & 15.1 & 3.5 & 0.23 & Y & Y \\ 
18 & 1.36e2 & 7.53e-2 & 2.01e-1 & 1.36e2 & 2.14e3 & 18.3 & 3.0 & 0.21 & N & N \\ 
19 & 9.44e2 & 1.21e0 & 2.05e0 & 9.41e2 & 4.41e3 & 15.0 & 2.0 & 0.24 & Y & Y \\ 
20 & 5.82e2 & 4.87e-1 & 9.52e-1 & 5.81e2 & 3.86e3 & 15.7 & 1.5 & 0.26 & N & N \\ 
21 & 1.30e3 & 1.22e0 & 2.70e0 & 1.30e3 & 2.47e3 & 18.7 & 3.0 & 0.26 & N & N \\ 
22 & 2.85e3 & 3.80e0 & 5.51e0 & 2.85e3 & 3.11e3 & 17.7 & 4.0 & 0.18 & Y & Y \\ 
23 & 2.44e3 & 2.55e0 & 4.46e0 & 2.44e3 & 6.28e3 & 15.7 & 4.0 & 0.23 & N & N  \\ 
24 & 1.39e3 & 1.34e0 & 2.77e0 & 1.39e3 & 6.85e3 & 15.5 & 2.0 & 0.20 & Y & Y \\ 
25 & 5.66e3 & 1.11e1 & 1.12e1 & 5.65e3 & 9.56e3 & 14.6 & 2.5 & 0.28 & Y & Y \\ 
26 & 4.06e3 & 6.26e0 & 8.48e0 & 4.05e3 & 6.31e3 & 15.8 & 3.0 & 0.34 & Y & Y \\
27 & 2.20e3 & 2.61e0 & 3.91e0 & 2.20e3 & 7.72e3 & 15.1 & 1.5 & 0.24 & Y & Y \\
28 & 3.87e2 & 5.72e-1 & 7.01e-1 & 3.86e2 & 2.07e3 & 14.5 & 2.0 & 0.26 & Y & Y \\
29 & 2.78e3 & 3.84e0 & 5.75e0 & 2.78e3 & 1.35e4 & 14.0 & 2.0 & 0.23 & Y & Y \\
30 & 4.81e3 & 9.36e0 & 9.32e0 & 4.81e3 & 5.76e3 & 15.5 & 3.0 & 0.28 & Y & Y \\
31 & 3.66e3 & 4.60e0 & 6.79e0 & 3.66e3 & 3.22e3 & 19.1 & 4.0 & 0.32 & N & N \\
32 & 3.42e2 & 3.93e0 & 3.78e-1 & 3.42e2 & 7.41e3 & 15.2 & 3.0 & 0.26 & N & N \\
33 & 1.68e3 & 3.84e0 & 2.77e0 & 1.67e3 & 4.42e3 & 16.7 & 4.0 & 0.40 & N & N \\
\hline
  \end{tabular*}
\end{center}
\end{table*}

\subsection{Hourglasses}\label{hourglasses}

Fig. \ref{hourglass} shows slices through and column density projections of the Model 
4 simulation in order to demonstrate the formation of hourglass field morphologies and
the strengthened magnetic fields. Model 4 was initialised with  
$\beta_{\rm plasma}=10.0$ resulting in a uniform magnetic field with magnitude of 
$0.363\,\mu$G. The radius of the initial cloud in code units was 1.0. In this magnetically 
weak case, the disc-like cloud has collapsed across the magnetic field lines under the 
influence of gravity. As can be seen in Fig. \ref{hourglass}a, after 46\,Myrs the cloud
has collapsed to a radius of $\sim0.2$ in code units. For comparison, the Model 3 cloud with an initial 
$\beta_{\rm plasma}=1.0$ has only shrunk slightly perpendicular to the magnetic field after a similar period
of evolution (as shown in Figs. \ref{evolution}e and \ref{evolution}f). The result of the collapse in Model 4
is still a disc-like cloud, but with a much smaller radius than the initial condition and an intensified magnetic field as shown in Fig.
\ref{hourglass}b. Across the denser regions of the cloud ($r<0.15$) the magnetic field
now averages 5 to 7$\,\mu$G. Some of the densest locations have field intensifications
up to $14\,\mu$G, approximately $40\times$ stronger than the initial field. Values of
$\beta_{\rm plasma}$ are around unity, corresponding to observational estimates in real molecular
clouds. This demonstrates that although the initial condition may not have a
realistic value of $\beta_{\rm plasma}$, the resulting molecular cloud
and dense condensations {\it do} have realistic properties.

\begin{figure*}
\centering
\includegraphics[width=140mm]{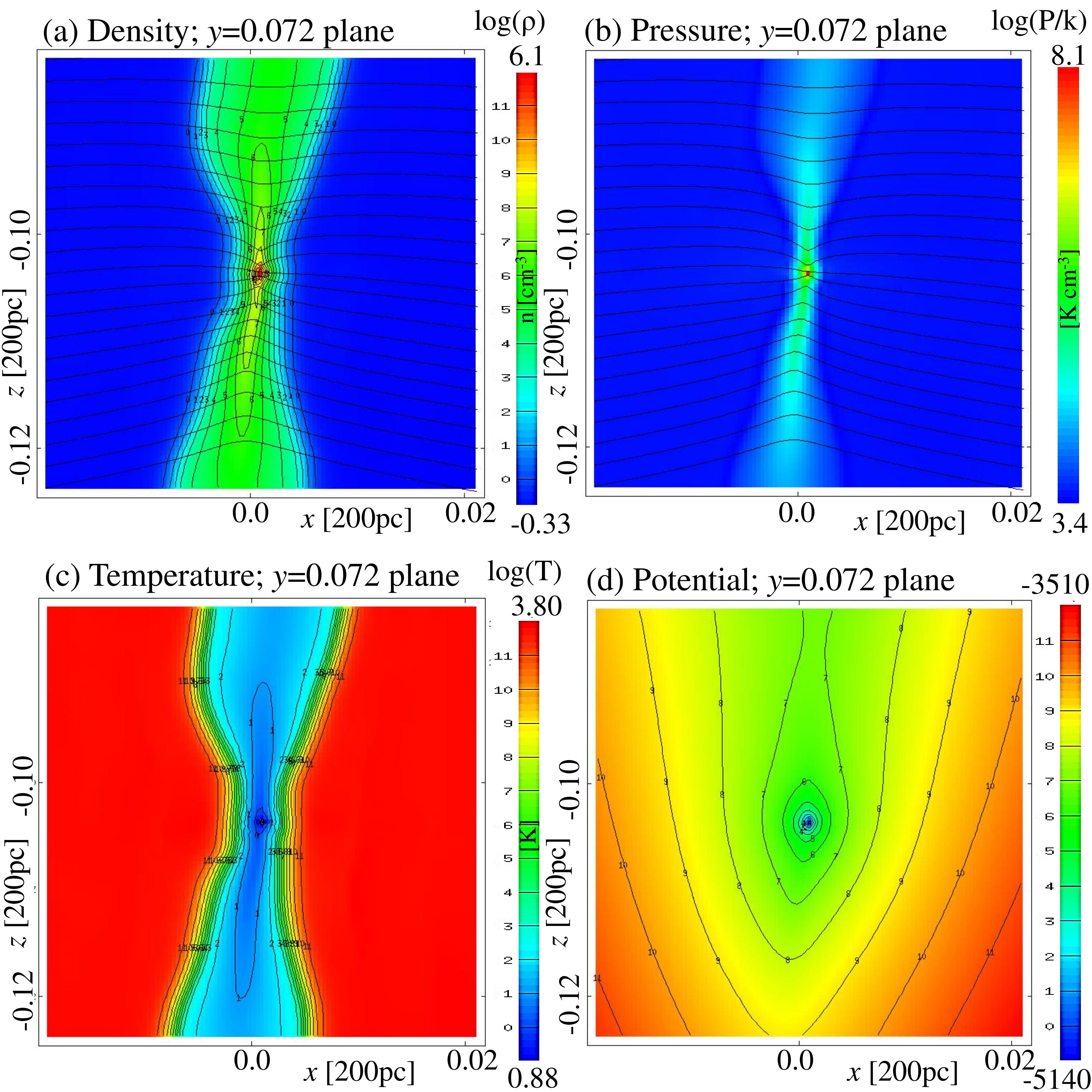}
\caption{Properties of Clump 5, the highest-density collapsing clump in the Model 5 simulation
at t=55.6\,Myrs. Shown are density, pressure, temperature and gravitational potential on the
plane at $y=0.072$, perpendicular to the sheet. Magnetic field lines are indicated on the density
and pressure plots. Raw data are available from https://doi.org/10.5518/897.}
\label{core1}
\end{figure*}

Fig. \ref{hourglass}c shows a column density projection at a slightly earlier time,
with magnetic field lines from the slice $y=0$ superimposed, as per the other panels in this
figure. In projection, this has the appearance of two dense clumps approaching 
each other, dragging the magnetic field lines with them. This is very reminiscent of 
the hourglass morphology of the magnetic field in the OMC1 region of the Orion A
filament where two clumps appear to be approaching each other and creating an 
hourglass magnetic field configuration \citep[see details in and fig. 1 of][]{pattle17}.
This demonstrates the
capability of the physical model to create intensified fields, with hourglass field
morphologies, that still preserve the large-scale ordered magnetic fields in the diffuse surroundings,
albeit as a remnant of the initial condition. Other models also
present field amplifications in the densest regions, e.g. similar values of the magnetic
field in the densest clump formed in Model 5 discussed in Sec. 4.5.1.
The hourglass magnetic field configuration compares well to observations,
for example in \cite{kandori17}, who derive the
magnetic field configuration around the starless dense core FeSt 1-457.

\subsection{Clumps}\label{coll}

We now consider the high-resolution Model 5 resimulation of the central section of Model 3 and 
identify any massive clumps that have formed. The FellWalker clump identification 
algorithm \citep{berry15} has been implemented into MG in order to do this. The implementation,
testing and application to the HD case is described in detail in Paper V.

\begin{figure*}
\centering
\includegraphics[width=140mm]{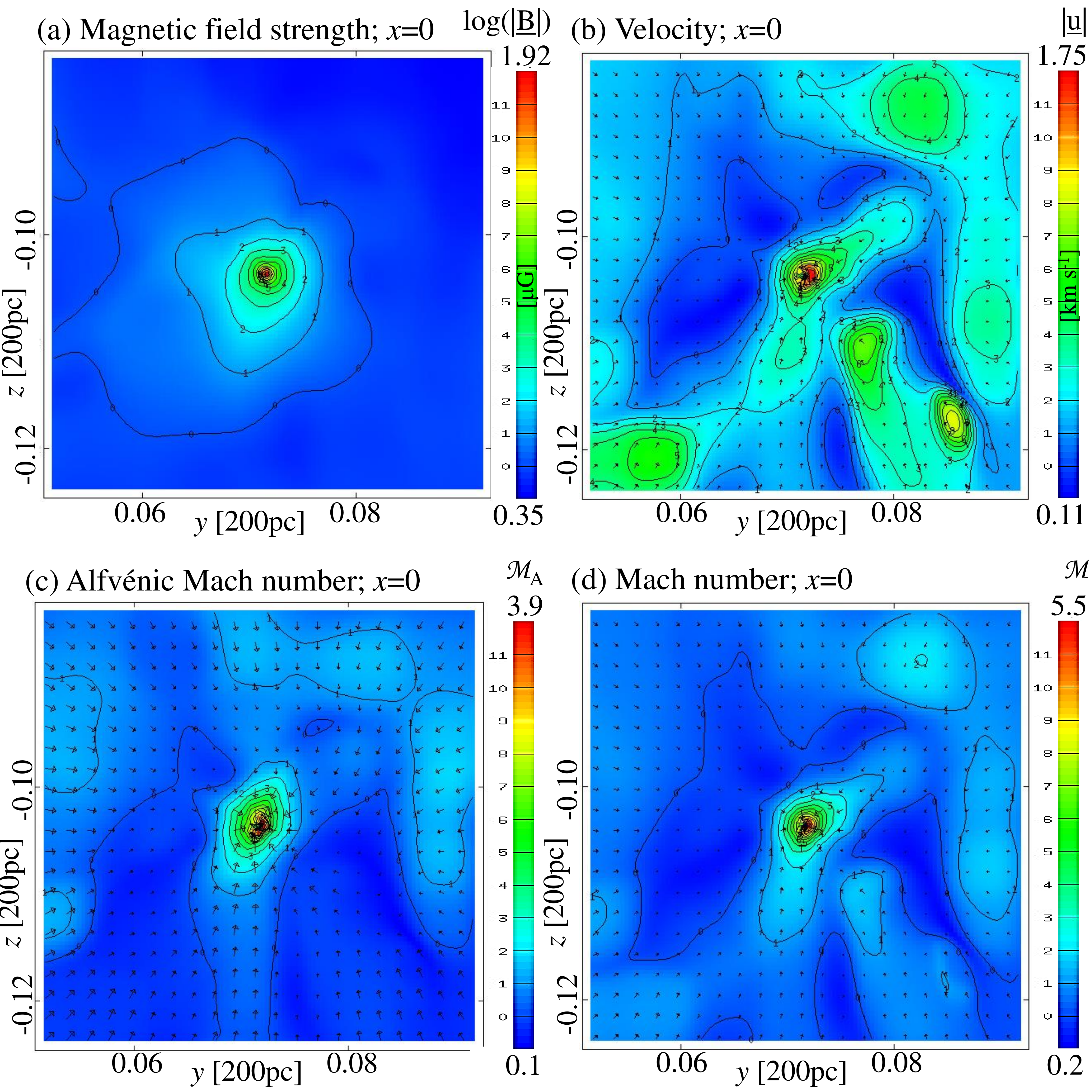}
\caption{Further properties of Clump 5, the highest-density collapsing clump in the Model 5 
simulation. Shown are magnetic field strength, velocity, Alfv\'enic Mach number and flow Mach 
number on the plane at $x=0$, parallel to the sheet. Velocity and Mach numbers are shown in 
the frame of the clump, not the simulation.
Raw data are available from https://doi.org/10.5518/897.}
\label{core2}
\end{figure*}

FellWalker has been applied to the Model 5 simulation after 5.5\,Myrs of evolution. This is in 
addition to the preceding 50.1\,Myrs of evolution of Model 3. The algorithm detected 33 clumps, 
detailed in Table \ref{tableclumps}, 20 of which are gravitationally unstable 
(E$_{\rm grav} >$ E$_{\rm th}$ + E$_{\rm kn}$ + E$_{\rm mag}$). At this time the highest density in the 
simulation, in the collapsing core of Clump 5, has reached the point at which it the resolution should be 
increased by adding other level of AMR. However, this is a good point to analyse the simulation 
since we have a reasonable number of gravitationally bound clumps
and the computational expense of extra AMR levels is very high. 

\cite{bergin07} review the properties of clumps based
on \cite{loren89} and \cite{williams94}: mass 50-500\,M$_{\odot}$; size 0.3-3\,pc; mean 
density 10$^3$-10$^4$\,cm$^{-3}$, velocity dispersion 0.3-3\,km\,s$^{-1}$; sound crossing 
time 1\,Myr; gas temperature 10-20\,K. These values fit those of the 33 clumps very well, 
except that the clumps are rather more massive. In each 
clump, the majority ($>95\%$) of the mass is in the cold phase. Minimum temperatures, which
occur in the inner regions of each clump where the lowest velocity dispersion also occurs, are in good 
agreement with Bergin \& Tafalla's review. Images of the clumps can be found in the accompanying
data archive at https://dx.doi.org/10.5518/897. The majority of the clumps are approximately spherical.

\subsubsection{Clump 5 - an individual clump with a collapsing core}

\begin{figure*}
\centering
\includegraphics[width=150mm]{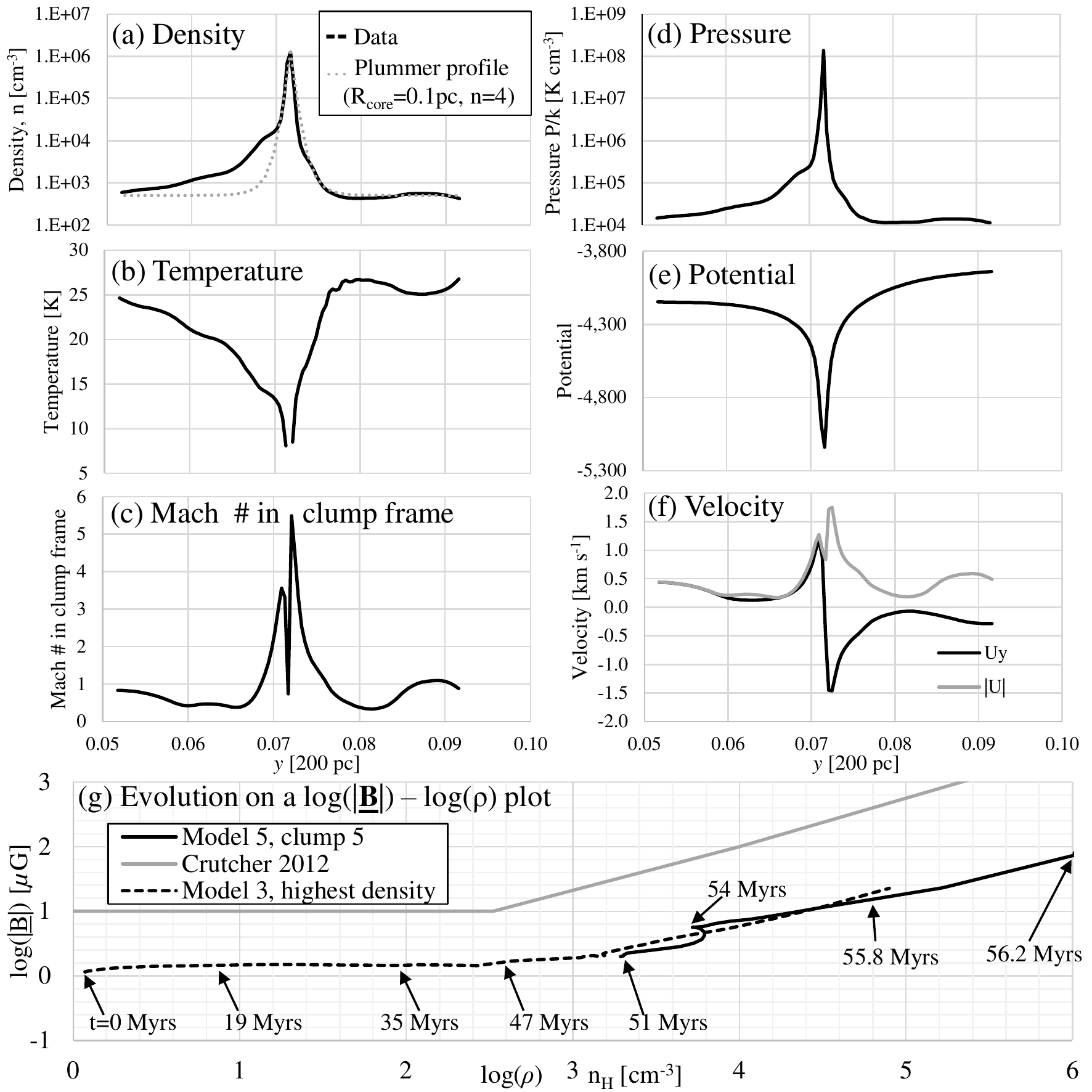}
\caption{Properties of the highest-density collapsing clump in the Model 5 simulation, \#5
in Table \ref{tableclumps}. Profiles are cuts along $y$ through the central position of the clump.
The bottom graph shows the track of the clump in magnetic field strength versus density
as it evolves with time. Raw data are available from https://doi.org/10.5518/897.}
\label{profiles}
\end{figure*}

Turning to examine an individual clump in more detail, Figs. \ref{core1} and \ref{core2} show 
slices through Clump 5 and its collapsing core. The slices and profiles are cut through the position of the gravitational 
potential minimum of the clump, at (x, y, z = 9.77e-04 (0.19\,pc),  7.17e-02 (14.3\,pc), -1.04e-01 (-20.8\,pc))
in the grid from -60\,pc to +60\,pc in all three directions.
Fig. \ref{core1}a shows the remarkably regular density profile described in the previous section, 
on a slice perpendicular to the sheet. This is common across the clumps in the Model 5 simulation and
different to the HD results in Paper V, where asymmetric and complex clumps were typical, formed
by the absorption of the filamentary network and clump collisions. Fig. \ref{core1}b shows
the high pressure in the collapsing core of the clump, a convincing signature of gravitational collapse.
Whilst Clump 5 and its collapsing core seem quantitatively different from the other
clumps in Table \ref{tableclumps}, it should be noted that this is merely the first of the clumps in the simulation to undergo
core-collapse. Others would do the same if the simulation was evolved further beyond this point.

Notable from the magnetic field lines shown in Figs \ref{core1}a and \ref{core1}b is the deformation
of the field in the local area of the clump, showing the effect on the field of the gravitational collapse.
Fig. \ref{core1}c shows the temperature on the slice through the clump, and the deep cold well
at the collapsing core of the clump, dropping to a realistic 8.1\,K. The potential, shown in Fig. 
\ref{core1}d is remarkable for its smoothness across the whole slice compared to the relative 
complexity apparent in density. In the context of simulated HD and MHD molecular clouds, 
gravitational potential is clearly useful for identification of distinct clumps with structure-finding 
tools such as FellWalker or CLUMPFIND. Fig. \ref{core2}a shows the magnetic field on a slice
now across (parallel to) the sheet and perpendicular to the initial background field,
indicating that the magnetic field has been very strongly intensified from an initial value of 
1.15\,$\mu$G up to nearly 100\,$\mu$G. This intensification and magnitude compare well to the
magnetic field observations made 
by the BISTRO collaboration \cite[e.g.][]{pattle17,liu19b,wang19,coude19,doi20}.
Similarly, the collapse under the influence of gravity has also led to large infall velocities, on the 
order of 2\,km\,s$^{-1}$. This is not an unreasonable velocity compared to observations, but 
the more important question concerns whether this velocity is supersonic and/or super-Alfv\'enic.
We find that it is both, as shown in Figs. \ref{core2}c and \ref{core2}d. Specifically, at the
centre, the infall velocity has an Alfv\'enic Mach number up to 4 and a thermal Mach number
up to 5.5. Such values appear to be comparable to observations and the kind of initial conditions
used by the turbulent star formation community to initialise such simulations. 

\cite{crutcher12} 
comments that the relative importance of turbulent to magnetic energy is addressed by a number 
of linear polarization results. He notes that the observed column density power spectrum in several 
cloud complexes is best reproduced by simulations that are super-Alfv\'enic, but agreement in the 
mean alignment of fields in cores and the surrounding medium cannot be reproduced by globally 
super-Alfv\'enic cloud models. This suggests that any simulation should
be sub-Alfv\'enic on the global scales, but transition to super-Alfv\'enic on the small scales of
gravitational collapse. Observations of the starless dense core FeSt 1-457, including ordered
field lines around the core, support the conclusion of transition from magnetic sub- to 
supercriticality at the boundary of the core \citep{kandori18}. This is a severe test of models 
and simulations, as noted by Crutcher, one which the model here clearly passes.

Fig. \ref{profiles} shows cuts along the $y$ axis through the position of the potential minimum 
of Clump 5. Clump 5 has a fairly uniform density distribution. Around the central peak, this is 
well fitted  by a Plummer-like density distribution over three orders of magnitude in density
(i.e. the classic Plummer-like profile introduced by \cite{whitworth01} with an observatinally
confined power-law index of 4). The fit takes a central density of n = 
1.23e6 cm$^{-3}$ from the data and a minimal central flat radius of $\approx$~0.1\,pc. This is 
not unreasonable given the FWHM of 0.2\,pc of the peak.
Panels (b) to (f) of Fig. \ref{profiles} show further detail of Clump 5, quantifying data 
presented on the slices in Figs. \ref{core1} and \ref{core2}.

Fig. \ref{profiles}g shows the evolution of the highest density location in the Model 3 simulation
and Clump 5 in the Model 5 simulation on a plot of magnetic field versus density. It is clear
that for the majority of the simulated time (the first 50\,Myrs), whilst the density increases,
the magnetic field at the highest density location remains approximately constant - replicating
the flat region of the \cite{crutcher12} fit. In fact the magnetic field only begins to intensify
when density reaches exactly the turning point fit by Crutcher, after approximately 47\,Myrs
of evolution. From that point onward, the magnetic field at the highest density location and
then at the position of Clump 5, grows with a remarkably similar power-law index to that of 
Crutcher's fit at high-density.  

\begin{figure}
\centering
\includegraphics[width=80mm]{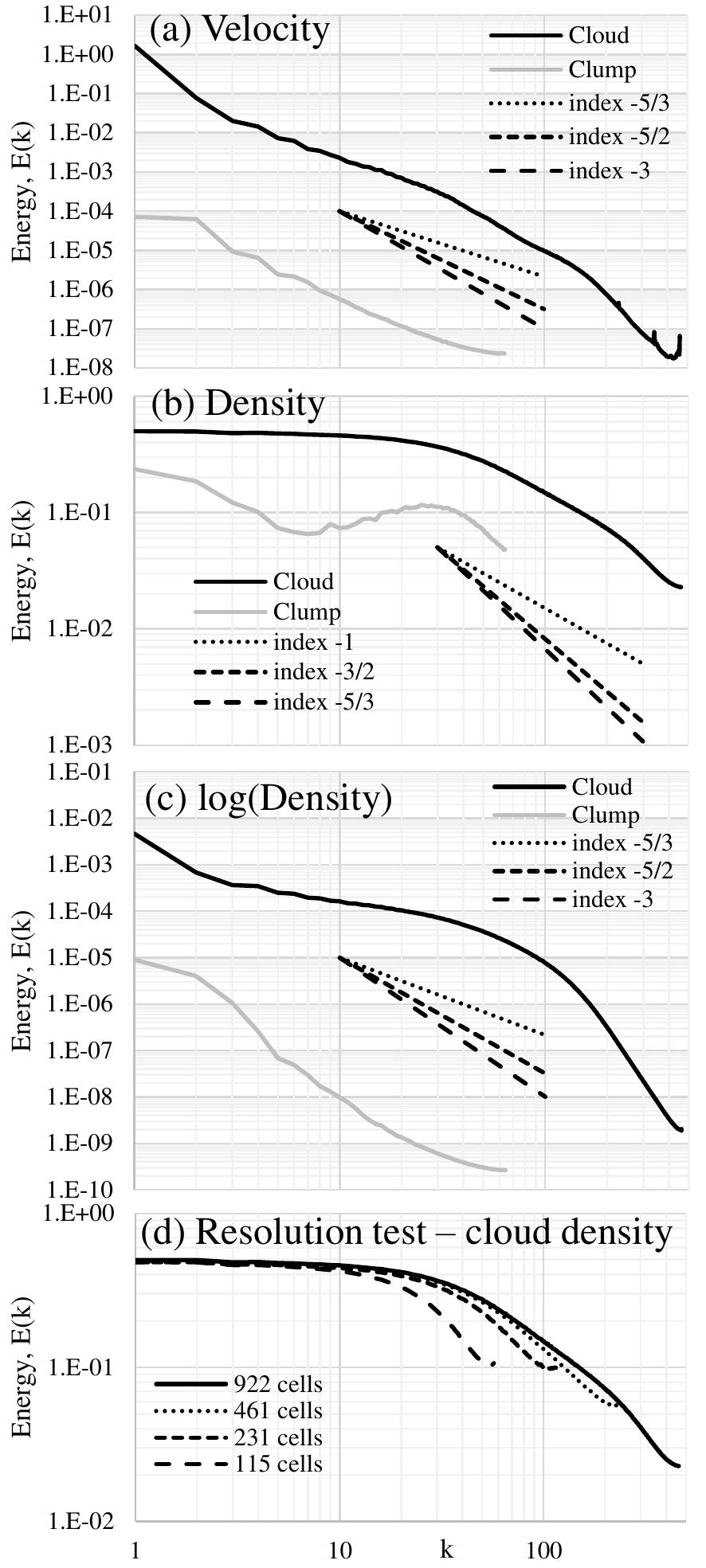}
\caption{Instantaneous snapshot power spectra of (a) velocity, (b) density 
and (c) log(density) for the cloud as a whole and for Clump 
5 at t=55.6\,Myr. Shown in (d) is the result of interpolating
data onto different levels of the AMR grid in order to demonstrate convergence.
Raw data are available from https://doi.org/10.5518/897.}
\label{spectra}
\end{figure}

\subsection{Power spectra}\label{spec}

In Fig. \ref{spectra}, we show  the snapshot power spectra of velocity
and  density for  both  the cloud  as  a  whole and  for  Clump 5,  at
t=55.6\,Myrs.   The power  spectra have  been calculated  in the  same
manner as in  Paper V, using the same tested  technique.  The velocity
spectra  has  been  calculated  from the  complete  velocity  3-vector
(v$_x$,  v$_y$, v$_z$).   As in  Paper V,  there is  no projection  or
smoothing of the vector into two components on a plane, as is known to
affect such spectra  \citep{medina14}. A uniform grid
  at the finest AMR level is used to generate the spectra, as was done
  but not specified  in Paper V.  In regions where  the finest grid is
  not present, this  was generated using the  projection operator from
  coarser grids.

The power spectra of velocity in Fig. \ref{spectra}a shows that both the cloud and Clump 5 display
an inertial range (region in wavenumber with slope of constant gradient) greater than one order 
of magnitude. The cube encompassing the cloud transformed for this analysis was 360\,pc on a 
side. Between wavenumbers 3 and 30 (or physical scales from 120\,pc to 12\,pc), the spectral
index is close to the Kolmogorov 5/3 spectrum generally observed in fully established turbulence
over an extended inertial range covering several decades in wavenumber. We do not have that
extent of inertial range, so this is not fully established turbulence, but this is very similar to the HD 
case in Paper V, albeit at smaller $k$ (larger physical scales). As noted in Paper V, this is also in 
good agreement with observationally derived velocity power spectra, for example, an index 
of $-1.81 \pm 0.10$ derived by \cite{padoan06} over a similarly short inertial range (little more than
one decade) for the Perseus molecular cloud complex. There is some indication of a spectral break 
in the cloud spectrum at $k=3$ (120\,pc), which indicates flow on the scale of the cloud. 
At $k=30$, there is a break, beyond which the spectrum steepens to an index around -3, 
corresponding roughly to the separation scale of structure driven by the thermal instability, similar 
again to Paper V.

The velocity spectra of Clump 5 is based on a cube 10\,pc on a side and so in a similar manner to 
Paper V, the power of the spectra is reduced by a factor 36$^3$ in order to allow for direct comparisons between
cloud and clump. In this case, the spectrum has a relatively steep initial index of -3,
in agreement with cloud spectra given the maximum scale of this clump spectra is 10\,pc and
indicative of consistent sampling of the same physical conditions. Such an index is also consistent
with that simulated by others on clump scales and below \citep{medina14}. The
spectra also tails upwards at the smallest scales. This may be indicative of the supersonic flows
at the smallest scales observed in Clump 5, but could as easily be an artefact of reaching the limits
of numerical resolution as such a curl upwards at highest $k$ is typical at the limit, as we have noted
before \citep{wareing09,wareing10}.

Although originally stated in paper V, it bears repeating here as it now also applies to the MHD case, 
that the stationary diffuse initial condition has generated large-scale flows with a turbulence-like 
-5/3 spectrum. The inertial range of the spectrum
goes from cloud to clump scales (120 to 12\,pc). The MHD simulations initially generate a remarkably 
laminar flow along the field lines, suggesting a transition towards turbulence as gravity takes over.
This clearly resembles the flow and field configuration observed with such instruments as {\it Planck},
as for example shown in Paper IV around the Rosette nebula. 

The power spectra of density are shown in Fig. \ref{spectra}b. The cloud spectrum is flat until
$k\sim30$ (12\,pc) and then breaks to an index of approximately -1 from $k=30$ to $k=200$,
corresponding to physical scales of 12 to 1.8\,pc. On first inspection, the flat low-$k$ spectrum 
may be surprising. It is worth bearing in mind that along the majority of the lines through the domain
with constant $y$ and $z$, parallel to the $x$ axis, the sheet is akin to a Delta function. The 
Fourier Transform of a Delta function is a constant. On the $x\sim0$ plane, the sheet is
approximately a top-hat function. The Fourier Transform of a top-hat function is a sinc
function. A three-dimensional Fourier Transform collapsed to one-dimension in order to obtain 
the power spectra is thus going to be some combination of constant power and then a steep
spectrum corresponding to the peaks of the sinc function. The density spectrum of the clump 
shows indications of the same curve up to $k=6$ as the cloud spectrum, albeit over a very small
range in $k$. Then there is a wide peak around $k=25$ to 30, corresponding to the physical
spherical scale of the clump potential well ($<$0.5\,pc). This peak is likely to be responsible for 
the upturn of the whole cloud spectrum at the largest value of $k\sim450$, as there is no
corresponding upturn of the Clump 5 spectra, suggesting the simulation is well-resolved in
density. The upturn would appear to be resolution independent, as shown in
Fig. \ref{spectra}d, adding credibility to a physical origin in the smallest scales of the simulation.

Power spectra of the logarithm of density have been shown
by \cite{kowal07} to exhibit a Kolmogorov-like behaviour when there is strong contrast of density.
Those authors note that the logarithmic operation significantly filters the extreme values of the 
density, stopping them from distorting the spectra. Fig. \ref{spectra}c shows such spectra here have a spectral index 
in the cloud around -1 at large scales, steepening at small scales. There is no clear inertial range for
the clump power spectra of the logarithm of density. It would appear that the high contrast between 
the dense sheet and the surroundings significantly affects the power spectrum of the clump, if less
so the spectrum of the cloud.
The flattening observed in density spectra is therefore due to both the sheet-like nature of the
cloud and, similarly to \cite{kowal07}, the dense small-scale structures generated across the sheet.

Apart from the size of the cloud, the only scale imposed by the initial conditions is that of 
the perturbation on the grid scale, which is 0.39\,pc for Model 3. As in our previous simulations, 
there is no evidence of this scale in the spectra.

\section{Summary and conclusions}\label{conclusions}

The purpose of the simulations described above was to determine whether thermal instability 
in a diffuse cloud could produce gravitationally collapsing objects, without any other influences 
(e.g. turbulence) or external disturbance (pressure wave, shock or collision).
Our previous work (Papers I-IV), at a lower resolution of $\geq$0.29\,pc, had revealed that clumpy clouds 
form in the hydrodynamic case and corrugated sheet-like clouds, that in projection appear
filamentary, form in the magnetic case. At high resolution, conclusive gravitational collapse
has been demonstrated in the purely hydrodynamic case (Paper V).
The suite of high-resolution simulations carried out here, with $<$0.4\,pc resolution,
up to 0.078\,pc high resolution, have now conclusively demonstrated that thermal
instability in a diffuse magnetic medium can generate cold and dense enough structure 
to allow self-gravity to take over and start the star formation process. The total
time scale for this to happen is on the order of 55\,Myrs, although the structure would
only be considered a molecular cloud for the previous 20\,Myrs. 

We have noted the following:-
\begin{enumerate}
\item Diffuse thermally unstable material flows sub-Alfv\'enically along field lines and
changes the spherical diffuse initial condition into a thick disc with voids and dense regions.
Eventually this sheet evolves to a single, thin sheet. In the magnetically
supercritical case, this sheet then collapses perpendicular to the field.
\item The relationship obtained by \cite{crutcher12} is reproduced in full, both following
the evolution of the whole cloud and that of a single clump. The turning point of Crutcher's
Bayesian fit at n $\sim300$\,cm$^{-3}$ is reproduced, as is the power-law gradient 
above that density. Agreement is shown between \cite{crutcher12}, observational data and 
Mach 10 turbulent simulations of other authors \citep{li15}.
\item Striation-like structure appears around the cloud and internally across the voids 
during the formation of the dense regions. These striations can have
column densities comparable to those observed around molecular clouds, but are in the
warm diffuse stable material, not cold, dense material as observed in CO emission. 
These diffuse striations are similar to Galactic fibres. They are not the same as the cold,
dense striations observed around molecular clouds in CO emission \citep[e.g.][]{goldsmith08}
and modelled by \cite{tritsis16}.
\item The sheet-like molecular cloud goes through a period of oscillation about the 
gravitational potential minimum as it settles to its final equilibrium. During this time, the
sheet resembles an integral-shaped structure, with breaks in the velocity structure along
the projected filamentary structure and associated red- and blue-shifted velocity patterns.
This is similar to several observational results \cite[e.g.][]{stutz16,lobos19,shi19}.
\item The gravity-dominated collapse of magnetically supercritical sheet-like clouds drags 
the magnetic field into an hourglass-like morphology and intensifies the magnetic field 
strength. The process creates field morphologies and strengths that resemble those observed 
\cite[e.g., see BISTRO collaboration results:][]{pattle17,liu19b,wang19,coude19,doi20}.
\item Application of the FellWalker algorithm \citep{berry15} to a high-resolution (0.078\,pc)
resimulation of a portion of Model 3 finds 33 clumps with properties similar to those deduced
from observations. Of these 33 clumps, 20 are confirmed to be gravitationally bound and 
collapsing. The densest clump has a density in its collapsing core (at the time at which
the simulation is stopped) six orders of magnitude greater than the initial condition. It 
is collapsing on the order of a realistic free-fall time. The clump has supersonic
and super-Alfv\'enic infall velocities, as opposed to sub-sonic and sub-Alfv\'enic velocities
across the cloud as a whole, in agreement with the observational characteristics deduced
by \cite{crutcher12}.
\item Velocity power spectra of the cloud as a whole and this densest clump 
show spectral indices that are turbulence-like (spectral index of $-5/3$) over a short inertial range 
(approximately one decade of wavenumber), even with the stationary initial diffuse condition.
This is the result of a large-scale laminar-like flow along the field lines, 
with structure on small scales.
\item Velocity and density power spectra resemble turbulent initial conditions implying 
that 1D power spectra only offer a limited tool to discern between models of star formation.
\item The most massive clumps are found to be Jeans unstable and therefore
undergo run-away gravitational collapse. Thermal instability, combined with self-gravity, 
is therefore able to produce collapsing clumps from a diffuse cloud, even in the presence
of a dynamically significant magnetic field.
\end{enumerate}

It should be noted that we have not included thermal conductivity, nor
fully resolved the Field length according to the conditions set by \cite{koyama04}.
Even so, starting from highly idealised initial conditions, we still obtain convergent 
simulations that accurately capture the large-scale structure of the resulting thermally 
bistable medium. A convergence study and a full discussion of why it is not necessary to 
include thermal conduction, nor resolve the Field length in this case, is presented in 
Appendix \ref{sectapp1}.

Immediate future work will now consider the effect of single star and cluster feedback in
both the Paper V HD results and the MHD results presented herein, by the inclusion
of a robust cluster-particle formation technique.

\section*{Acknowledgments}

We acknowledge support from the Science and Technology Facilities
Council (STFC, Research Grant ST/P00041X/1). The calculations 
herein were performed on
the DiRAC 1 Facility at Leeds jointly funded by STFC, the Large
Facilities Capital Fund of BIS and the University of Leeds and on
other facilities at the University of Leeds. 
We thank David Hughes at Leeds for the provision of IDL scripts which
formed the basis of the power spectra analysis presented in this work.

\section*{Data Availability}

The data underlying this article are available in the Research Data Leeds
Repository, at https://doi.org/10.5518/897.

\appendix

\section{Thermal Conductivity}\label{sectapp1}

\cite{field65} showed that, in the  absence of thermal conduction, the
growth rate of  the condensation mode of thermal  instability does not
have  a maximum,  but  increases to  a finite  positive  limit as  the
wavelength  tends  to zero.   However,  thermal  conduction induces  a
maximum in the growth rate  and also stabilises modes whose wavelength
is smaller than the Field length

\begin{equation}
\lambda_F = 2 \pi \left[ \frac{\kappa  T}{\rho ( \rho L_\rho - T L_T)}
  \right]^{1/2}
\label{field}
\end{equation}
\noindent
Here $\kappa$ is the thermal  conductivity and $L_\rho$, $L_T$ are the
derivatives of the energy loss rate per unit mass, $L$, w.r.t. density
and  temperature.   Note  that  this  is  only  defined  for  isobaric
instability, $\rho  L_\rho -  T L_T >  0$, and  $\lambda_F \rightarrow
\infty$ at the boundaries of the unstable region.

\cite{koyama04} use
\begin{equation}
\lambda_F = \left( \frac{T \kappa}{\rho L_c} \right)^{1/2},
\label{koyamafield}
\end{equation}
\noindent
where $L_c$ is the cooling rate  per unit mass. This is very different
from equation (\ref{field}): it is  significantly smaller and does not
tend to infinity at the boundaries of the unstable region.  Fig.  3 in
\cite{falle20} shows  a comparison  between these two  definitions for
our energy  loss rate (\citealt{koyama02}) and  a thermal conductivity
given by
\begin{equation}
  \kappa = 2.5 \times 10^3 T^{1/2},
\label{kappa}
\end{equation}
\noindent
(\citealt{parker53}).  This  is  due  to  neutrals  and  is  therefore
unaffected by the magnetic field. Note  that there is an error in this
figure: equation (\ref{koyamafield}) is multiplied by $2 \pi$.

\cite{koyama04} argue  that simulations of thermal  instability do not
converge  unless one  includes  thermal conductivity  and resolves  the
Field length. Formally, this is  true for linear perturbations, but at
our initial  density, $n =  1.1$, the growth  rate has a  rather broad
maximum  at $\lambda  = 8.95$  pc  (\citealt{falle20}),  so there is no
wavelength that  is particularly  favoured in  the linear  regime. The
maximum growth  rate is $0.1774$  Myr$^{-1}$, which is quite  close to
the zero  conductivity limit of $0.1788$  Myr$^{-1}$.  \cite{koyama04}
used a somewhat higher density and a different energy loss function so
that the maximum might be somewhat sharper in their case.

\begin{figure*}
\centering
\includegraphics[width=160mm]{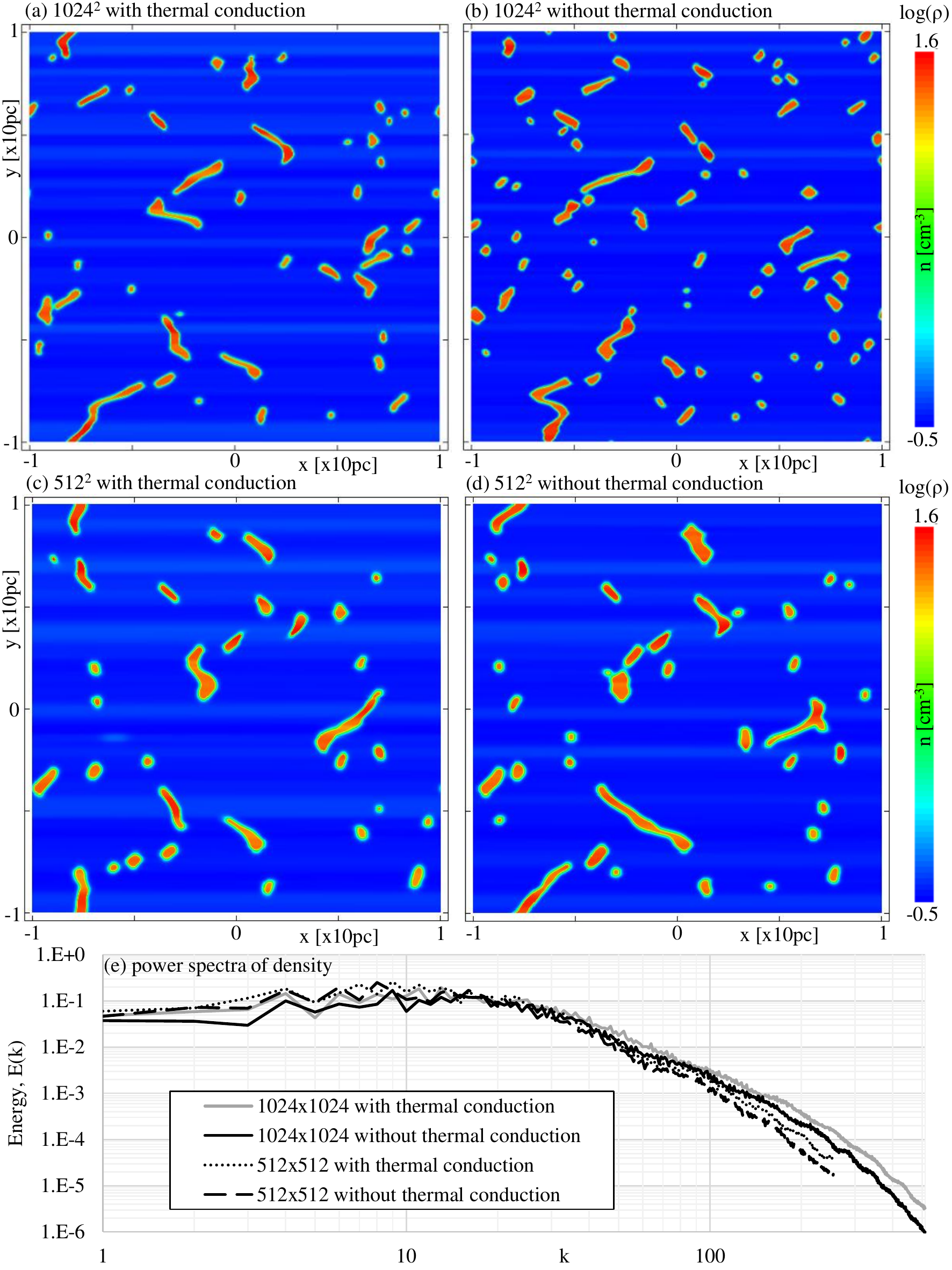}
\caption{A comparison of simulations with and without thermal conduction
and at different resolutions, as discussed in the text.
Raw data are available from https://doi.org/10.5518/897.}
\label{app1}
\end{figure*}

Fig.   \ref{app1} shows  a  comparison between  calculations with  and
without  conduction.   These  are two  dimensional  calculations  with
periodic  boundary  conditions  with   random  initial  conditions  as
described in Section  2. These initial conditions were imposed on the
fully refined $512^2$ grid and projected
onto the fully refined $1024^2$ grid, thereby ensuring the same  
wavelengths are present on both grids. This
is as close as we can get to the same initial conditions for both
resolutions. The alternative of imposing the initial conditions on the 
1024$^2$ grid and projecting to 512$^2$ is less satisfactory since 
it introduces wavelengths on the $1024^2$ grid which cannot be 
represented on the $512^2$ grid.
At the unperturbed initial  density, $n = 1.1$, equation
(\ref{field})  gives  $\lambda_F =  0.564$  pc.   There are  therefore
approximately $14$ and  $29$ cells in the initial Field  length at the
low and  high resolutions, which is an adequate representation of the Field scale.  Note
that equation (\ref{koyamafield}) gives $\lambda_F = 0.0587$ pc.

It can be seen that these calculations all give very
  similar  results  and are  much  the  same  as the  two  dimensional
  calculations  in  \cite{wareing16}  which  had a  much  larger  mesh
  spacing  of  $0.156$  pc.  The   main  difference  is  that  thermal
  conduction  reduces the  number  of  small clouds,  which  is to  be
  expected  and agrees  with \cite{hennebelle07a}  who found  that the
  mass  distribution of  the larger  clouds  is not  much affected  by
  thermal conduction.   In our calculations  there is a collapse  to a
  corrugated sheet, during which such small clouds are mostly absorbed
  by the larger ones.
It should be noted that whilst there are slight differences
between the power spectra at large and small-scales, they are all converged at 
intermediate wavenumbers, $k=11 - 40$, further increasing confidence
in the thermal structure that forms at these size-scales. The factor of 4 difference
which is apparent at the smallest scales (though inverted and less at the largest
scales) is not down to differences in the initial condition. This could be
down to slight differences in growth rates at the finest cell sizes.

Other authors have found similar results e.g.  \cite{piontek04} used a
larger value of the thermal  conductivity than (\ref{kappa}) but argue
that this  does not have  much effect on  the final properties  of the
larger clouds.   \cite{gazol05} did not include  thermal conduction in
their simulations with driven turbulence, but they also found that the
small scale structure did not have  a significant effect on the global
properties.  \cite{inoue15}  included thermal  conduction, 
but were  not able to  resolve the Field  length in their  large scale
calculations.  However,   they  also   found  that   the
properties of the thermally bistable medium converged on large scales,
because  ``most  of the  mass  of  the  cold  gas created  by  thermal
instability is contained in large clumps that are formed by the growth
of large-scale fluctuations."

AMR cannot be used for the initial evolution of the
instability since it would derefine unless the error tolerance is very
small. This makes  it impossible to resolve the Field  length in large
scale calculations such  as these since  the minimum Field
length is $0.0485$ pc  at $n = 6.2517$ (note that  the Field length is
only meaningful  in the unstable  region).  It is  therefore fortunate
that  the final  distribution of  large clouds  is insensitive  to the
value of the thermal conductivity.

\label{lastpage}

\end{document}